\documentclass[prd,showpacs,floatfix,amsmath,amssymb,floatfix]{revtex4}
\usepackage{amsfonts}
\usepackage{mathrsfs}
\usepackage{amssymb}
\usepackage{amsmath}
\usepackage{pifont}
\usepackage{extarrows}
\usepackage{slashed}
\usepackage{curves}
\usepackage{color}
\usepackage{lscape}
\usepackage{multirow}
\usepackage{makecell}
\usepackage{indentfirst}
\usepackage{graphicx}
\allowdisplaybreaks

\begin{document}

\title{The electromagnetic decays of the charmed and bottom baryons in chiral perturbation theory}

\author{Nan Jiang}\email{Jiangn@pku.edu.cn}
\author{Xiao-Lin Chen}\email{chenxl@pku.edu.cn}
\author{Shi-Lin Zhu}\email{zhusl@pku.edu.cn}

\affiliation {$^1$School of Physics and State Key Laboratory of
Nuclear Physics and Technology, Peking University, Beijing 100871, China\\
$^2$Collaborative Innovation Center of Quantum Matter, Beijing
100871, China}



\begin{abstract}

We have investigated the electromagnetic decays of the antitriplet
and sextet charmed baryon systems with $J^P= \frac{1}{2}^+,
\frac{3}{2}^+$ in the framework of the heavy baryon chiral
perturbation theory. We first construct the chiral Lagrangians at
$O(p^2)$ and $O(p^3)$. Then we calculate the electromagnetic (EM)
decay amplitudes of the charmed baryon systems up to $O(p^3)$. With
the help of the quark model, we estimate the low energy constants.
The numerical results of the EM decay widths show good convergence
of the chiral expansion. We notice that the two neutral EM decay
processes $\Xi_c'^0\rightarrow\gamma+\Xi_c^0$ and
${\Xi_c^*}'^0\rightarrow\gamma+\Xi_c^0$ are strongly suppressed by
the SU(3) U-spin flavor symmetry. With the same formalism, we also
estimate the EM decay widths of the bottomed baryons. The EM decay
widths of the heavy baryons may be measured at facilities such as
LHCb and JPARC. The explicit chiral structures of the heavy baryon
decay amplitudes derived in this work may be useful to the possible
chiral extrapolations of the future lattice simulations of these EM
decay amplitudes.

\end{abstract}

\maketitle

\section{Introduction}\label{sec1}

The low energy behavior of quantum chromodynamics (QCD) is highly
nonperturbative and complicated. The study of the hadron properties
helps us explore the infrared structure of QCD. In the past decades
many heavy baryons were discovered experimentally
\cite{Agashe:2014kda,Wick:2011jn,Aaltonen:2011sf}.

The heavy flavored baryon systems containing a single heavy quark
are particularly interesting. In the heavy quark limit, there exists
the additional heavy quark spin and flavor symmetry when the heavy
quark mass goes to infinity. The strong dynamics within the heavy
baryon is governed by the two light quarks while the heavy quark
acts as a color source and the spectator. The physical observables
can be expanded in terms of $1/m_Q$ where $m_Q$ is the heavy quark
mass.

In this work, we focus on the heavy baryons containing a single
charm or bottom quark and investigate their electromagnetic (EM)
decays. The heavy baryon ground states contain the
spin-$\frac{1}{2}$ sextet, antitriplet and the spin-$\frac{3}{2}$
sextet. For brevity, we refer to them as the 6, $\bar3$ and $6^*$
dimensional baryons in the following. We will explore the property
of the electromagnetic transitions between the heavy baryon
multiplets.

In literature, various theoretical approaches were employed to study
the magnetic moments and EM decays of the light baryons
\cite{Uppal:1992cc,Majethiya:2009vx,Puglia:2000jy,Meissner:1997hn,Geng:2009hh,Zhu:1998ih,
Cheng:1992xi,Cheng:1993kp,Cheung:1993ux,Bahtiyar:2015sga,Banuls:1999br,Korner:1994nh,
Cho:1994vg,Savage:1994wa,Ivanov:1998wj,Ivanov:1999bk,Tawfiq:1999cf}.
We have performed a systematic study of the chiral corrections to
the heavy baryon mass and axial charge in Ref. \cite{Jiang:2014ena}.
In this work, we will investigate the electromagnetic decay width of
the antitriplet and sextet heavy baryon systems in the framework of
heavy baryon chiral perturbation theory (HBChPT). We truncate the
chiral expansion at $O(p^3)$. The tree-level electromagnetic decay
amplitude appears at $O(p^2)$ and $O(p^3)$. We include the chiral
loop corrections up to $O(p^3)$. The recoil corrections in terms of
$\frac{1}{M}$ appear at $O(p^4)$ or even high order. As will be seen
below, our results show good convergence of the chiral expansion.

There are twelve possible EM decay channels. Because of the
constraint of flavor structure, only ten channels arise at the
leading order. At $O(p^3)$, there appear six additional EM
interaction low energy constants involving the 6-dimensional and
$6^*$-dimensional baryons at the tree level. In contrast, the loop
contributions appear at $O(p^3)$ and contribute to all twelve
channels.

The EM width is proportional to the cubic of the photon momentum,
which is roughly equivalent to the mass difference of the initial
and final states. The average value of the mass difference is 127
MeV between the $\bar3$ and 6-dimensional charmed baryons, 67 MeV
between the 6 and $6^*$-dimensional baryons, 194 MeV between $\bar3$
and $6^*$-dimensional baryons. In the calculation of the EM decay
widths, we use the experimental values of the charmed baryon masses
as input to derive the photon momentum. The EM decay widths of
different channels differ greatly, mainly due to the large
difference of the photon momentum. In the appendix, we also extend
the same formalism to estimate the EM widths of the bottomed
baryons.

This paper is organized as follows. In Sec. II, we introduce the
effective chiral Lagrangians up to $O(p^3)$. In Sec. III, we
calculate the EM decay amplitude and width order by order. In Sec.
IV, we determine the low energy constants (LECs) using the quark
model. In Sec. V, we discuss the numerical results and summarize. We
collect some useful formulae and the numerical results of the EM
decay of the bottom baryons in the appendix.

\section{The Chiral Lagrangians}\label{sec2}

\subsection{The Leading Order Chiral Lagrangians}\label{sec2.1}

The three light quarks $u, d, s$ form the fundamental representation
under the flavor SU(3) transformation. The heavy baryons can be
classified into multiplets in flavor SU(3) space. Now we concentrate
on the ground states of the charmed baryon. When the two light
quarks are in the flavor antitriplet, the quantum number of the
heavy baryon is $J^P=\frac{1}{2}^+$. When the two light quarks are
in the symmetric flavor sextet, the quantum number of the heavy
baryon can be either $J^P=\frac{1}{2}^+$ or $\frac{3}{2}^+$.

In this work, we investigate the heavy baryon systems employing the
Chiral Perturbation Theory (ChPT) \cite{Weinberg:1978kz}, which
provides a systematic expansion of the physical observables in terms
of small momentum $p$ and the mass of Goldstone bosons $m$. ChPT has
been widely used to study the low energy hadron interaction. In the
early stage, it was employed to study the purely mesonic system
\cite{Gasser:1983yg,Gasser:1984gg}. Later it was extended to discuss
the baryon-meson system
\cite{Gasser:1987rb,Jenkins:1990jv,Bernard:1992qa,Bernard:1992nc}.
In order to deal with the heavy baryon system, the heavy baryon
chiral perturbation theory (HBChPT) was developed
\cite{Bernard:1996gq,Mojzis:1997tu,Fettes:1998ud,Fettes:2001cr},
which provides a convenient framework to make a dual expansion in
terms of both the small momentum and $\frac{1}{M}$, where $M$ is the
heavy baryon mass.

In the flavor space, the pseudoscalar meson fields and
spin-$\frac{1}{2}$ baryon multiplets are defined as follows
\[\phi=\left(
  \begin{array}{ccc}
  \pi^0+\frac{1}{\sqrt{3}}\eta&\sqrt{2}\pi^+&\sqrt{2}K^+\\
  \sqrt{2}\pi^-&-\pi^0+\frac{1}{\sqrt{3}}\eta&\sqrt{2}K^0\\
  \sqrt{2}K^-&\sqrt{2}\bar{K}^0&-\frac{2}{\sqrt{3}}\eta\\
  \end{array}
\right)\]
\begin{eqnarray}\label{eq3}
B_{\bar{3}}=\left(
                \begin{array}{ccc}
                0&\Lambda_c^+&\Xi_c^+\\
                -\Lambda_c^+&0&\Xi_c^0\\
                -\Xi_c^+&-\Xi_c^0&0\\
                \end{array}
              \right),\quad
B_{6}=\left(
                \begin{array}{ccc}
                \Sigma_c^{++}&\frac{1}{\sqrt{2}}\Sigma_c^+&\frac{1}{\sqrt{2}}\Xi_c'^+\\
                \frac{1}{\sqrt{2}}\Sigma_c^+&\Sigma_c^0&\frac{1}{\sqrt{2}}\Xi_c'^0\\
                \frac{1}{\sqrt{2}}\Xi_c'^+&\frac{1}{\sqrt{2}}\Xi_c'^0&\Omega_c^0\\
                \end{array}
              \right)
\end{eqnarray}
The spin-$\frac{3}{2}$ baryons $B_6^{*\mu}$ are the so-called
Rarita-Schwinger vector-spinor fields \cite{Johnson:1960vt}, which
are similar to $B_6$. In the framework of ChPT, one can construct
the Lagrangians order by order while the specific building blocks
satisfying the chiral transformation are defined as follows
\cite{Scherer:2002tk,Bernard:1995dp}.
\begin{eqnarray}\label{eq4}
U(x)&=&e^{\frac{i}{F_0}\phi(x)},\quad u^2=U\nonumber\\
\Gamma_\mu&=&\frac{1}{2}(u^\dagger\partial_\mu u+u\partial_\mu u^\dagger)\nonumber\\
u_\mu&=&\frac{i}{2}(u^\dagger\partial_\mu u-u\partial_\mu u^\dagger)\nonumber\\
D_\mu B&=&\partial_\mu B+\Gamma_\mu B+B\Gamma_\mu^T\nonumber\\
\nabla_\mu U&=&\partial_\mu U-ir_\mu U+iUl_\mu
\end{eqnarray}
\begin{equation}
r_\mu=l_\mu=-eQ\mathcal{A}_\mu,\quad Q=\left(
                                 \begin{array}{ccc}
                                 \frac{2}{3}&0&0\\
                                 0&-\frac{1}{3}&0\\
                                 0&0&-\frac{1}{3}\\
                                 \end{array}
                               \right)
\end{equation}
The superscript $T$ denotes the transpose in the flavor space. The
pion decay constant $F_0\approx92.4$ MeV. The pure meson Lagrangian
in the lowest order $O(p^2)$ is
\begin{equation}\label{eq5}
\mathcal{L}^{(2)}=\frac{F_0^2}{4}\textrm{Tr}[\nabla_\mu U(\nabla_\mu
U)^\dagger]
\end{equation}
The leading order pseudoscalar meson and heavy baryon Lagrangians at
$O(p^1)$ read \cite{Yan:1992gz}
\begin{eqnarray}
\mathcal{L}_0^{(1)}&=&\frac{1}{2}\textrm{Tr}[\bar{B}_{\bar{3}}(i\slashed D-M_{\bar{3}})B_{\bar{3}}]+\textrm{Tr}[\bar{B}_6(i\slashed D-M_6)B_6]\nonumber\\
&&+\textrm{Tr}\{\bar{B}_6^{*\mu}[-g_{\mu\nu}(i\slashed D-M_6^*)+i(\gamma_\mu D_\nu+\gamma_\nu D_\mu)-\gamma_\mu(i\slashed D+M_6^*)\gamma_\nu]B_6^{*\nu}\}\label{eq6}\\
\mathcal{L}_{\textrm{int}}^{(1)}&=&g_1\textrm{Tr}(\bar{B}_6\slashed u\gamma_5B_6)+g_2[\textrm{Tr}(\bar{B}_6\slashed u\gamma_5B_{\bar{3}})+\textrm{h.c.}]+g_3[\textrm{Tr}(\bar{B}_6^{*\mu}u_\mu B_6)+\textrm{h.c.}]\nonumber\\
&&+g_4[\textrm{Tr}(\bar{B}_6^{*\mu}u_\mu
B_{\bar{3}})+\textrm{h.c.}]+g_5\textrm{Tr}(\bar{B}_6^{*\nu}\slashed
u\gamma_5B_{6\nu}^*)+g_6\textrm{Tr}(\bar{B}_{\bar{3}}\slashed
u\gamma_5B_{\bar{3}})\label{eq7}
\end{eqnarray}
where $\mathcal{L}_0^{(1)}$ and $\mathcal{L}_{\textrm{int}}^{(1)}$
are the free and interaction part. From the quark model and flavor
SU(3) symmetry, the axial coupling constants $g_6=0$, $g_1=0.98$,
$g_1=-\sqrt{\frac{8}{3}}g_2$ \cite{Yan:1992gz}. The heavy quark spin
flavor symmetry leads to the following relations among these
coupling constants, i.e.,
$g_3=\frac{\sqrt{3}}{2}g_1,g_5=-\frac{3}{2}g_1,g_4=-\sqrt{3}g_2$.
Numerically, $g_1=0.98,g_2=-0.60,g_5=-1.47,g_3=0.85,g_4=1.04$.

In the framework of HBChPT, the baryon field $B$ is decomposed into
the large component $\mathcal{N}$ and the small component
$\mathcal{H}$. Their relationships are
\begin{eqnarray*}
&&B=e^{-iMv\cdot x}(\mathcal{N}+\mathcal{H})\\
&&\mathcal{N}=e^{iMv\cdot x}\frac{1+\slashed
v}{2}B,\quad\mathcal{H}=e^{iMv\cdot x}\frac{1-\slashed v}{2}B,
\end{eqnarray*}
where $v^\mu$ is the on shell velocity. For the spin-$\frac{3}{2}$
baryon, the large component is denoted as $\mathcal{T}^\mu$. Such a
decomposition is equivalent to separating the off-shell momentum $k$
from the whole momentum $p$ of the heavy baryon while the on shell
part is $Mv$. The reduced effective Lagrangian only relies on the
large component. Omitting some details, the nonrelativistic
Lagrangian in the heavy quark limit reads
\begin{eqnarray}\label{eq8}
\hat{\mathcal{L}}^{(1)}&=&\frac{1}{2}\textrm{Tr}[\bar{\mathcal{N}}_{\bar{3}}(iv\cdot D)\mathcal{N}_{\bar{3}}]+\textrm{Tr}[\bar{\mathcal{N}}_6(iv\cdot D-\delta_1)\mathcal{N}_6]+\textrm{Tr}\{\bar{\mathcal{T}}^\rho[-g_{\rho\sigma}(iv\cdot D-\delta_2)]\mathcal{T}^\sigma\}\nonumber\\
&&+2g_1\textrm{Tr}(\bar{\mathcal{N}}_6S\cdot u\mathcal{N}_6)+2g_2[\textrm{Tr}(\bar{\mathcal{N}}_6S\cdot u\mathcal{N}_{\bar{3}})+\textrm{h.c.}]+g_3[\textrm{Tr}(\bar{\mathcal{T}}^\mu u_\mu \mathcal{N}_6)+\textrm{h.c.}]\nonumber\\
&&+g_4[\textrm{Tr}(\bar{\mathcal{T}}^\mu u_\mu
\mathcal{N}_{\bar{3}})+\textrm{h.c.}]+2g_5\textrm{Tr}(\bar{\mathcal{T}}^\nu
S\cdot
u\mathcal{T}_\nu)+2g_6\textrm{Tr}(\bar{\mathcal{N}}_{\bar{3}}S\cdot
u\mathcal{N}_{\bar{3}})
\end{eqnarray}
The mass difference parameters are defined as
$\delta_1=M_6-M_{\bar3}, \delta_2=M_{6^*}-M_6$. In the isospin
symmetry limit, $\delta_1=126.52$ MeV, $\delta_2=67.03$ MeV. We have
not kept the $\frac{1}{M}$ corrections to the Lagrangian because
their contributions are of higher order.

\subsection{The higher order Lagrangians for the EM decays}\label{sec2.2}

Considering the conservation of the electric charge and energy,
there exist twelve electromagnetic decay channels, which can be
classified into three types:

$B_6\rightarrow B_{\bar3}+\gamma:$
\[
\left\{
\begin{aligned}
&\Sigma_c^+\rightarrow\Lambda_c^++\gamma\\
&\Xi_c'^+\rightarrow\Xi_c^++\gamma\\
&\Xi_c'^0\rightarrow\Xi_c^0+\gamma
\end{aligned}
\right.
\]

$B_6^*\rightarrow B_6+\gamma:$
\[
\left\{
\begin{aligned}
&\Sigma_c^{*++}\rightarrow\Sigma_c^{++}+\gamma\\
&\Sigma_c^{*+}\rightarrow\Sigma_c^{+}+\gamma\\
&\Sigma_c^{*0}\rightarrow\Sigma_c^{0}+\gamma\\
&{\Xi_c^*}'^+\rightarrow\Xi_c'^++\gamma\\
&{\Xi_c^*}'^0\rightarrow\Xi_c'^0+\gamma\\
&\Omega_c^{*0}\rightarrow\Omega_c^0+\gamma
\end{aligned}
\right.
\]

$B_6^*\rightarrow B_{\bar3}+\gamma:$
\[
\left\{
\begin{aligned}
&\Sigma_c^{*+}\rightarrow\Lambda_c^++\gamma\\
&{\Xi_c^*}'^+\rightarrow\Xi_c^++\gamma\\
&{\Xi_c^*}'^0\rightarrow\Xi_c^0+\gamma
\end{aligned}
\right.
\]

The leading order Lagrangians in Eqs. (\ref{eq6})-(\ref{eq7}) do not
contribute to the above EM decays. So the lowest order EM decay
diagrams arise from the $O(p^2)$ Lagrangians at the tree level. If
we consider the next leading order corrections, we need the
contribution from both the tree-level $O(p^3)$ Lagrangians and the
chiral loop corrections from the leading order Lagrangians.

The higher ordered Lagrangian can be constructed with the help of
the new building blocks $f^\pm_{\mu\nu}$.
\begin{eqnarray}
f^\pm_{\mu\nu}&=&u^\dagger f^R_{\mu\nu}u\pm uf^L_{\mu\nu}u^\dagger\\
f^R_{\mu\nu}&=&\partial_\mu r_\nu-\partial_\nu r_\mu-i[r_\mu,r_\nu]\\
f^L_{\mu\nu}&=&\partial_\mu l_\nu-\partial_\nu l_\mu-i[l_\mu,l_\nu]
\end{eqnarray}
\begin{equation}
r_\mu=l_\mu=-eQ_B\mathcal{A}_\mu,\quad Q_B=\left(
                                 \begin{array}{ccc}
                                 2&0&0\\
                                 0&0&0\\
                                 0&0&0\\
                                 \end{array}
                               \right)\textrm{for the charmed baryons},
                               \quad Q_B=\left(
                                 \begin{array}{ccc}
                                 1&0&0\\
                                 0&-1&0\\
                                 0&0&-1\\
                                 \end{array}
                               \right)\textrm{for the bottomed baryons}
\end{equation}

The transformation property of the building blocks can be seen in
Ref. \cite{Fettes:2000gb}. What we need in the following is that the
operator $f^\pm_{\mu\nu}$ transforms as the adjoint representation.
We notice that the direct product of $3 \times 6=8+10$ contains only
one adjoint representation. Therefore, there is one independent
interaction term in the $O(p^2)$ Lagrangians for the EM transitions
between the $\bar3$ and $6/6^*$-dimensional baryons.

In contrast, the direct product of $\bar 6 \times 6$ contains one
singlet and one octet. Accordingly, there are two independent
interaction terms for the EM transitions between the 6 and
$6^*$-dimensional baryons. The octet piece corresponds to the
$f^+_{\mu\nu}$ term while the singlet piece corresponds to the trace
of $f^+_{\mu\nu}$ in Eq.(\ref{eq13}). The $O(p^2)$ Lagrangians are
as follows
\begin{eqnarray}\label{eq13}
\left\{
\begin{aligned}
&\mathcal{L}^{(2)}_{\bar36}=\frac{f_2}{\Lambda_\chi}\textrm{Tr}\bar B_{\bar3}\sigma^{\mu\nu}f^+_{\mu\nu}B_{6}+\textrm{h.c.}\\
&\mathcal{L}^{(2)}_{66^*}=\frac{f_3}{\Lambda_\chi^2}\textrm{Tr}\bar B_{6}\epsilon^{\mu\nu\rho\sigma}f^+_{\mu\nu}iD_\rho B^{6^*}_\sigma+\frac{\tilde{f}_3}{\Lambda_\chi^2}\textrm{Tr}\bar B_{6}\epsilon^{\mu\nu\rho\sigma}\langle f^+_{\mu\nu}\rangle iD_\rho B^{6^*}_\sigma+\textrm{h.c.}\\
&\mathcal{L}^{(2)}_{\bar36^*}=\frac{f_4}{\Lambda_\chi^2}\textrm{Tr}\bar
B_{\bar3}\epsilon^{\mu\nu\rho\sigma}f^+_{\mu\nu}iD_\rho
B^{6^*}_\sigma+\textrm{h.c.}
\end{aligned}
\right.
\end{eqnarray}
where $\textrm{Tr}$ and $\langle\rangle$ denote the trace in the
flavor space, and $\Lambda_\chi=4\pi F_0$. The coefficients $f_i$'s
are the new dimensionless LECs. The 8-dimensional operator
$f^+_{\mu\nu}$ and the $\bar3,6$-dimensional baryon fields couple
into the possible flavor singlet.

Similarly, the $O(p^3)$ Lagrangians for the EM decays are
constructed as
\begin{eqnarray}\label{eq14}
\mathcal{L}^{(3)}_{66^*}&=&\frac{h_1}{\Lambda_\chi^3}\textrm{Tr}\bar B_6\nabla_\lambda f^+_{\mu\nu}\gamma^\lambda\gamma_5iD^{[\mu}B_{6^*}^{\nu]}+\frac{h_2}{\Lambda_\chi^3}\textrm{Tr}\bar B_6\nabla_\lambda f^+_{\mu\nu}\gamma^{[\mu}\gamma_5iD^{\nu]}B_{6^*}^{\lambda}+\frac{h_3}{\Lambda_\chi^3}\textrm{Tr}\bar B_6\nabla_\lambda f^+_{\mu\nu}iD^{\lambda}\gamma^{[\mu}\gamma_5B_{6^*}^{\nu]}+\textrm{h.c.}\\
&&+\frac{\tilde{h}_1}{\Lambda_\chi^3}\textrm{Tr}\bar
B_6\nabla_\lambda \langle
f^+_{\mu\nu}\rangle\gamma^\lambda\gamma_5iD^{[\mu}B_{6^*}^{\nu]}+\frac{\tilde{h}_2}{\Lambda_\chi^3}\textrm{Tr}\bar
B_6\nabla_\lambda \langle
f^+_{\mu\nu}\rangle\gamma^{[\mu}\gamma_5iD^{\nu]}B_{6^*}^{\lambda}+\frac{\tilde{h}_3}{\Lambda_\chi^3}\textrm{Tr}\bar
B_6\nabla_\lambda \langle f^+_{\mu\nu}\rangle
iD^{\lambda}\gamma^{[\mu}\gamma_5B_{6^*}^{\nu]}+\textrm{h.c.}\nonumber
\end{eqnarray}
\begin{eqnarray}\label{eq15}
\mathcal{L}^{(3)}_{\bar36^*}&=&\frac{h_4}{\Lambda_\chi^3}\textrm{Tr}\bar
B_{\bar3}\nabla_\lambda
f^+_{\mu\nu}\gamma^\lambda\gamma_5iD^{[\mu}B_{6^*}^{\nu]}+\frac{h_5}{\Lambda_\chi^3}\textrm{Tr}\bar
B_{\bar3}\nabla_\lambda
f^+_{\mu\nu}\gamma^{[\mu}\gamma_5iD^{\nu]}B_{6^*}^{\lambda}+\frac{h_6}{\Lambda_\chi^3}\textrm{Tr}\bar
B_{\bar3}\nabla_\lambda
f^+_{\mu\nu}iD^{\lambda}\gamma^{[\mu}\gamma_5B_{6^*}^{\nu]}+\textrm{h.c.}
\end{eqnarray}
where $D^{[\mu}B^{\nu]}=D^{\mu}B^{\nu}-D^{\nu}B^{\mu}$ and the other
commutators are similar. The coefficients $h_i$'s are also
dimensionless LECs.

Because of the constraint of the above flavor structure, we notice
that there does not exist EM transitions between the neutral $\bar3$
and $6/6^*$-dimensional baryons at the tree level up to $O(p^3)$.
This seemingly surprising result is the manifestation of the famous
flavor SU(3) U-spin symmetry, which interchanges the strange and
down quark. Notice that the s and d quarks carry the same charge.
With the exact SU(3) flavor symmetry, the s and d quarks are
identical particles both in QCD and QED! Therefore, the Pauli
principle forbids the s and d pair within the $\Xi_c^0$ baryon to
form the flavor antitriplet. In other words, the decay processes
$\Xi_c'^0\rightarrow\Xi_c^0+\gamma$ and
${\Xi_c^*}'^0\rightarrow\Xi_c^0+\gamma$ are forbidden at the tree
level up to $O(p^3)$.

If we consider the explicit SU(3) symmetry breaking, we can
construct the tree-level Lagrangians at $O(p^4)$ such as
\begin{eqnarray}\label{eq16}
\mathcal{L}^{(4)}\propto\textrm{Tr}\bar
B_{\bar3}\chi_+\sigma^{\mu\nu}\langle f^+_{\mu\nu}\rangle
B_{6}+\textrm{h.c.}
\end{eqnarray}
where $\chi_+$ is the building block relevant to the current quark
mass \cite{Fettes:2000gb}. Since we truncate the chiral expansion at
$O(p^3)$, we do not try to exhaust all the possible $O(p^4)$
Lagrangians.

On the other hand, the decay processes
$\Xi_c'^0\rightarrow\Xi_c^0+\gamma$ and
${\Xi_c^*}'^0\rightarrow\Xi_c^0+\gamma$ can occur through the chiral
loops at $O(p^3)$ if we consider the explicit SU(3) symmetry
breaking. Its divergence would be absorbed by the above tree-level
Lagrangians at $O(p^4)$. We will include these corrections carefully
in the following section.

\section{The EM Decay Widths in HBChPT}\label{sec3}

\subsection{The EM Decay Widths at $O(p^2)$}\label{sec3.1}

\begin{figure}[h]
\begin{center}
\begin{minipage}[h]{0.2\linewidth}
\begin{center}
  \includegraphics{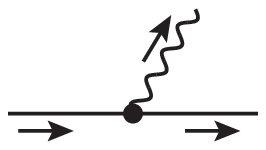}
\end{center}
\end{minipage}
\begin{minipage}[h]{0.2\linewidth}
\begin{center}
  \includegraphics{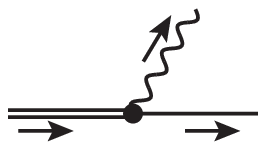}
\end{center}
\end{minipage}
\end{center}
  \caption{The $O(p^2)$ tree level process}\label{Fig.1}
\end{figure}

We calculate the EM decay amplitude in Coulomb gauge. For a
nonpolarized process, the summation of the photon polarizationyields
the metric $h_{\mu\nu}$
\[\sum_{r=1,2}\epsilon^{r*}_\mu\epsilon^r_\nu=-h_{\mu\nu}\]
We adopt the metric $h_{\mu\nu}$ in Ref. \cite{Pokorski:1987ed}.
\[h_{\mu\nu}=g_{\mu\nu}-\frac{l\cdot n(l_\mu n_\nu+l_\nu n_\mu)-l_\mu l_\nu}{(l\cdot n)^2-l^2}\]
where $n=(1,0,0,0)$ and $l$ is the photon 4-momentum. We plot the
lowest order Feynman diagrams from Eq.(\ref{eq13}) in Fig.
\ref{Fig.1}.
The single line represents a spin-$\frac{1}{2}$ baryon and double
line a spin-$\frac{3}{2}$ baryon. The solid dot represents an
$O(p^2)$ EM vertex.

Summing the photon polarization $r$, the spin directions of the
initial and final baryon states $S_A,S_B$, we obtain the square of
EM decay amplitude modular for the spin-$\frac{1}{2}$ to
$\frac{1}{2}$ and spin-$\frac{3}{2}$ to $\frac{1}{2}$ transitions
respectively.
\begin{eqnarray}
\sum_{r,S_A,S_B}\left|\mathcal{M}^{(2)}_{\frac{1}{2}-\frac{1}{2}}\right|^2=16e^2\left|D^{(2)}_{ij}\right|^2M_AM_B\ell^2
\end{eqnarray}
\begin{eqnarray}
\sum_{r,S_A,S_B}\left|\mathcal{M}^{(2)}_{\frac{1}{2}-\frac{3}{2}}\right|^2=\frac{128}{3}e^2\left|D^{(2)}_{ij}\right|^2M_AM_B^3\ell^2
\end{eqnarray}
where $M_A,M_B$ represent the baryon mass of the initial and the
final states. $\ell$ is the modular of the photon 3-momentum. In the
rest frame of the initial state, $\ell$ only depends on the baryon
mass.
\begin{eqnarray}\label{eq19}
\ell=\frac{(M_A+M_B)(M_B-M_A)}{2M_B}
\end{eqnarray}
For each channel, the coefficients $D^{(2)}_{ij}$'s are listed in
the 2nd column of Table \ref{CoefficientD}.

Since some LECs always appear in a combined form in the decay
amplitudes, we abbreviate $h_{ij}=h_i+h_j$ where $h_i$ etc arises
from the tree-level Lagrangians at $O(p^3)$. We also collect the
LECs $h_{ij}=h_i+h_j$ in Table \ref{CoefficientD}. For the bottomed
baryon systems, the LECs $f^b,h^b$ are marked with the superscript
$b$ for distinction. The corresponding coefficients are collected in
Table \ref{CoefficientD-b}.

\begin{table}[!h]
  \centering
  \caption{The coefficients $D^{(2)}_{ij},D^{(3)}_{ij,1},D^{(3)}_{ij,2}$ in the EM decay amplitudes.}\label{CoefficientD}
\begin{tabular}{cccc}
\Xhline{1.2pt}
EM decay processes & Coeff.$D^{(2)}_{ij}$ of $O(p^2)$ & Coeff.$D^{(3)}_{ij,1}$ of $O(p^3)$ & Coeff.$D^{(3)}_{ij,2}$ of $O(p^3)$\\
\Xhline{1.2pt}
$\Sigma_c^+\rightarrow\gamma+\Lambda_c^+$      &$-\frac{4\sqrt{2}f_2}{\Lambda_{\chi}}$& -  & -\\
$\Xi_c'^+\rightarrow\gamma+\Xi_c^+$            &$-\frac{4\sqrt{2}f_2}{\Lambda_{\chi}}$& -  & -\\
$\Xi_c'^0\rightarrow\gamma+\Xi_c^0$            &  -   & -  & -\\
\Xhline{1.2pt}
$\Sigma_c^{*++}\rightarrow\gamma+\Sigma_c^{++}$&$-\frac{4(f_3+\tilde{f}_3)}{\Lambda_{\chi}^2}$&$\frac{16(h_{13}+\tilde{h}_{13})}{\Lambda _{\chi }^3}$&$-\frac{16(h_{23}+\tilde{h}_{23})}{\Lambda _{\chi }^3}$\\
$\Sigma_c^{*+}\rightarrow\gamma+\Sigma_c^{+}$  &$-\frac{2(f_3+2\tilde{f}_3)}{\Lambda_{\chi}^2}$&$\frac{8(h_{13}+\tilde{h}_{13})}{\Lambda _{\chi }^3}$&$-\frac{16(h_{23}+\tilde{h}_{23})}{\Lambda _{\chi }^3}$\\
$\Sigma_c^{*0}\rightarrow\gamma+\Sigma_c^{0}$  &$-\frac{4\tilde{f}_3}{\Lambda_{\chi}^2}$&$\frac{16\tilde{h}_{13}}{\Lambda _{\chi }^3}$&$-\frac{16\tilde{h}_{23}}{\Lambda _{\chi }^3}$\\
${\Xi_c^*}'^+\rightarrow\gamma+\Xi_c'^+$       &$-\frac{2(f_3+2\tilde{f}_3)}{\Lambda_{\chi}^2}$&$\frac{8(h_{13}+\tilde{h}_{13})}{\Lambda _{\chi }^3}$&$-\frac{16(h_{23}+\tilde{h}_{23})}{\Lambda _{\chi }^3}$\\
${\Xi_c^*}'^0\rightarrow\gamma+\Xi_c'^0$       &$-\frac{4\tilde{f}_3}{\Lambda_{\chi}^2}$&$\frac{16\tilde{h}_{13}}{\Lambda _{\chi }^3}$&$-\frac{16\tilde{h}_{23}}{\Lambda _{\chi }^3}$\\
$\Omega_c^{*0}\rightarrow\gamma+\Omega_c^0$    &$-\frac{4\tilde{f}_3}{\Lambda_{\chi}^2}$&$\frac{16\tilde{h}_{13}}{\Lambda _{\chi }^3}$&$-\frac{16\tilde{h}_{23}}{\Lambda _{\chi }^3}$\\
\Xhline{1.2pt}
$\Sigma_c^{*+}\rightarrow\gamma+\Lambda_c^+$   &$-\frac{2\sqrt{2}f_4}{\Lambda_{\chi}^2}$&$\frac{8\sqrt{2}\tilde{h}_{46}}{\Lambda _{\chi }^3}$&$-\frac{8\sqrt{2}\tilde{h}_{56}}{\Lambda _{\chi }^3}$\\
${\Xi_c^*}'^+\rightarrow\gamma+\Xi_c^+$        &$-\frac{2\sqrt{2}f_4}{\Lambda_{\chi}^2}$&$\frac{8\sqrt{2}\tilde{h}_{46}}{\Lambda _{\chi }^3}$&$-\frac{8\sqrt{2}\tilde{h}_{56}}{\Lambda _{\chi }^3}$\\
${\Xi_c^*}'^0\rightarrow\gamma+\Xi_c^0$        & -    &-&-\\
\Xhline{1.2pt}
\end{tabular}
\end{table}

\begin{table}[h]
  \centering
  \caption{The coefficients $D^{(2)}_{ij},D^{(3)}_{ij,1},D^{(3)}_{ij,2}$ for the bottomed baryons.}\label{CoefficientD-b}
\begin{tabular}{cccc}
\Xhline{1.2pt}
EM decay processes & Coeff.$D^{(2)}_{ij}$ of $O(p^2)$ & Coeff.$D^{(3)}_{ij,1}$ of $O(p^3)$ & Coeff.$D^{(3)}_{ij,2}$ of $O(p^3)$\\
\Xhline{1.2pt}
$\Sigma_b^0\rightarrow\gamma+\Lambda_b^0$      &$-\frac{4\sqrt{2}f^b_2}{\Lambda_{\chi}}$& -  & -\\
$\Xi_b'^0\rightarrow\gamma+\Xi_b^0$            &$-\frac{4\sqrt{2}f^b_2}{\Lambda_{\chi}}$& -  & -\\
$\Xi_b'^-\rightarrow\gamma+\Xi_b^-$            &  -   & -  & -\\
\Xhline{1.2pt}
$\Sigma_b^{*+}\rightarrow\gamma+\Sigma_b^+$&$\frac{2(-f^b_3+\tilde{f}^b_3)}{\Lambda_{\chi}^2}$&$-\frac{8(-h^b_{13}+\tilde{h}^b_{13})}{\Lambda _{\chi }^3}$&$\frac{8(-h^b_{23}+\tilde{h}^b_{23})}{\Lambda _{\chi }^3}$\\
$\Sigma_b^{*0}\rightarrow\gamma+\Sigma_b^0$  &$\frac{2\tilde{f}^b_3}{\Lambda_{\chi}^2}$&$-\frac{8\tilde{h}^b_{13}}{\Lambda _{\chi }^3}$&$\frac{8\tilde{h}^b_{23}}{\Lambda _{\chi }^3}$\\
$\Sigma_b^{*-}\rightarrow\gamma+\Sigma_b^-$  &$\frac{2(f^b_3+\tilde{f}^b_3)}{\Lambda_{\chi}^2}$&$-\frac{8(h^b_{13}+\tilde{h}^b_{13})}{\Lambda _{\chi }^3}$&$\frac{8(h^b_{23}+\tilde{h}^b_{23})}{\Lambda _{\chi }^3}$\\
${\Xi_b^*}'^0\rightarrow\gamma+\Xi_b'^0$       &$\frac{2\tilde{f}^b_3}{\Lambda_{\chi}^2}$&$-\frac{8\tilde{h}^b_{13}}{\Lambda _{\chi }^3}$&$\frac{8\tilde{h}^b_{23}}{\Lambda _{\chi }^3}$\\
${\Xi_b^*}'^-\rightarrow\gamma+\Xi_b'^-$       &$\frac{2(f^b_3+\tilde{f}^b_3)}{\Lambda_{\chi}^2}$&$-\frac{8(h^b_{13}+\tilde{h}^b_{13})}{\Lambda _{\chi }^3}$&$\frac{8(h_{13}+\tilde{h}_{13})}{\Lambda _{\chi }^3}$\\
$\Omega_b^{*-}\rightarrow\gamma+\Omega_b^-$    &$\frac{2(f^b_3+\tilde{f}^b_3)}{\Lambda_{\chi}^2}$&$-\frac{8(h^b_{13}+\tilde{h}^b_{13})}{\Lambda _{\chi }^3}$&$\frac{8(h^b_{13}+\tilde{h}^b_{13})}{\Lambda _{\chi }^3}$\\
\Xhline{1.2pt}
$\Sigma_b^{*0}\rightarrow\gamma+\Lambda_b^0$   &$-\frac{2\sqrt{2}f^b_4}{\Lambda_{\chi}^2}$&$\frac{8\sqrt{2}\tilde{h}^b_{46}}{\Lambda _{\chi }^3}$&$-\frac{8\sqrt{2}\tilde{h}^b_{56}}{\Lambda _{\chi }^3}$\\
${\Xi_b^*}'^0\rightarrow\gamma+\Xi_b^0$        &$-\frac{2\sqrt{2}f^b_4}{\Lambda_{\chi}^2}$&$\frac{8\sqrt{2}\tilde{h}^b_{46}}{\Lambda _{\chi }^3}$&$-\frac{8\sqrt{2}\tilde{h}^b_{56}}{\Lambda _{\chi }^3}$\\
${\Xi_b^*}'^-\rightarrow\gamma+\Xi_b^-$        & -    &-&-\\
\Xhline{1.2pt}
\end{tabular}
\end{table}

\subsection{The $O(p^3)$ contributions at the tree level}\label{sec4.1}

\begin{figure}[h]
\begin{center}
  \includegraphics{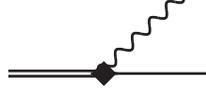}
\end{center}
  \caption{The $O(p^3)$ tree level process.}\label{Fig.2}
\end{figure}

The next leading order contributions arise from the tree-level
Lagrangians in Eqs.(\ref{eq14},\ref{eq15}) and chiral loop
corrections from Eqs.(\ref{eq5})-(\ref{eq7}). The Lagrangians in
Eqs.(\ref{eq13},\ref{eq14}) contribute to the tree-level process
between the spin-$\frac{1}{2}$ and $\frac{3}{2}$ baryons in Fig.
\ref{Fig.2}. In fact, there are two other Lagrangians involving
$f^+_{\mu\nu}$ at $O(p^3)$. One is $\bar B\gamma^\mu\nabla^\nu
f^+_{\mu\nu}B$ which leads to the EM decay amplitude with the factor
$l_\mu l^\nu\epsilon_\nu-l^2\epsilon_\mu$. Under the Coulomb gauge,
this term vanishes. The other possible interaction Lagrangian is
$\bar B\gamma_5f^+_{\mu\nu}D_\rho B_\sigma$. In the leading order of
the heavy baryon expansion, its contribution also vanishes since
$\bar{\mathcal{N}}\gamma_5\mathcal{T_\sigma}=0$.

There are two Lorentz structures from Fig. \ref{Fig.2}. To sum up,
all the EM decay amplitudes at the tree level up to $O(p^3)$ can be
written as
\begin{eqnarray}
\left\{
\begin{aligned}
&i\mathcal{M}^{(2)}=D^{(2)}M_B(l_\mu\epsilon_\nu-l_\nu\epsilon_\mu)\epsilon^{\mu\nu\rho\sigma}v_\rho\bar u^{S_A}u^{S_B}_\sigma\\
&i\mathcal{M}^{(3)}=iD^{(3)}_1M_Bv\cdot l\bar
u^{S_A}S^{\mu'}l_{\mu'}\epsilon_{\sigma'}u^{S_B}_{\sigma'}+iD^{(3)}_2M_Bv\cdot
l\bar u^{S_A}S^{\mu'}\epsilon_{\mu'}l_{\sigma'}u^{S_B}_{\sigma'}
\end{aligned}
\right.
\end{eqnarray}
where the Pauli-Lubanski spin operator $S^\mu$ is defined as
$S^\mu=-\frac{1}{2}\gamma_5(\gamma^\mu\slashed v-v^\mu)$. The
coefficients $D^{(3)}_1,D^{(3)}_2$'s are also listed in Table
\ref{CoefficientD}. In Coulomb gauge, the cross term between the
$O(p^2)$ and $O(p^3)$ EM decay amplitudes reads
\begin{eqnarray}
\sum_{r,S_A,S_B}(i\mathcal{M}^{(2)})^*i\mathcal{M}^{(3)}+(i\mathcal{M}^{(3)})^*i\mathcal{M}^{(2)}=-\frac{32}{3}M_AM_B^3\ell^3D^{(2)}(D^{(3)}_1-D^{(3)}_2)
\end{eqnarray}
The square of the decay amplitude from the $O(p^3)$ diagram is
\begin{eqnarray}
\sum_{r,S_A,S_B}(|i\mathcal{M}_1|^2+|i\mathcal{M}_2|^2)=\frac{8}{3}\left(\left|D^{(3)}_{ij,1}\right|^2+\left|D^{(3)}_{ij,2}\right|^2\right)M_AM_B^3\ell^4
\end{eqnarray}

\subsection{The $O(p^3)$ chiral loop contributions}\label{sec3.3}

\begin{figure}[h]
\begin{center}
  \includegraphics[scale=0.8]{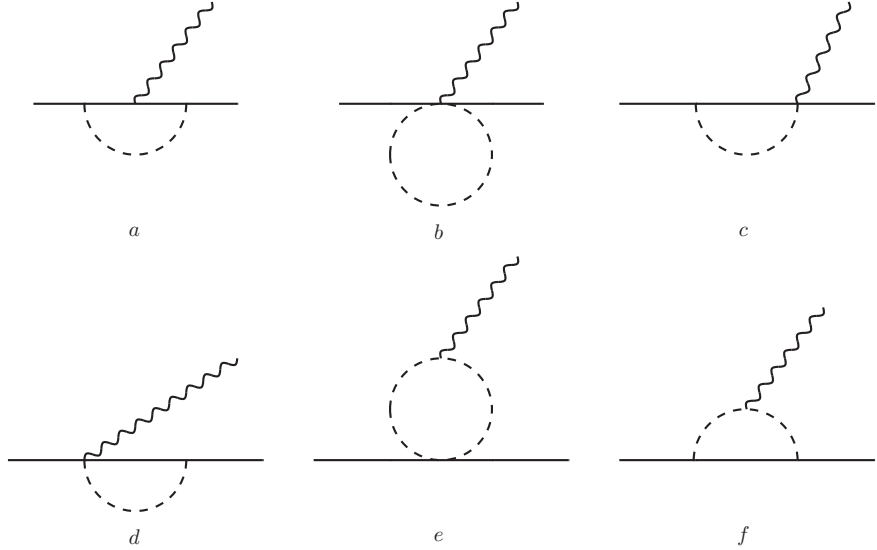}
\end{center}
  \caption{$O(p^3)$ loop level process}\label{Fig.3}
\end{figure}

At the one-loop level, there are six possible Feynman diagrams as
shown in Fig. \ref{Fig.3}. All the vertices are from Eqs.
(\ref{eq5},\ref{eq6},\ref{eq7}). In diagram a, the photon vertex is
from the tree-level EM interaction term while the meson vertex from
the strong interaction terms. In diagram b, the photon-meson vertex
is from the chiral connection. In diagram c and d, both two vertices
are from the strong interaction and seagull terms. In diagram e, the
meson vertex arises from the chiral connection.

The external baryons in diagrams b and e must be the same. So they
do not contribute to the EM decay process. Diagrams c and d vanish
in the heavy baryon mass limit. The nontrivial recoil corrections
appear at $O(p^4)$ which is beyond our chiral truncation order in
this work. The decay amplitude from diagram a between spin
$\frac{3}{2}$ to $\frac{1}{2}$ baryons is proportional to either
$S\cdot v=0$ or $v\cdot\mathcal{T}=0$. Hence its contribution
vanishes. Similarly the decay amplitude between spin $\frac{1}{2}$
to $\frac{1}{2}$ baryons from diagram a is always proportional to
$\epsilon\cdot v$. Under Coulomb gauge, this amplitude also
vanishes.

\begin{figure}[h]
\begin{center}
  \includegraphics[scale=0.8]{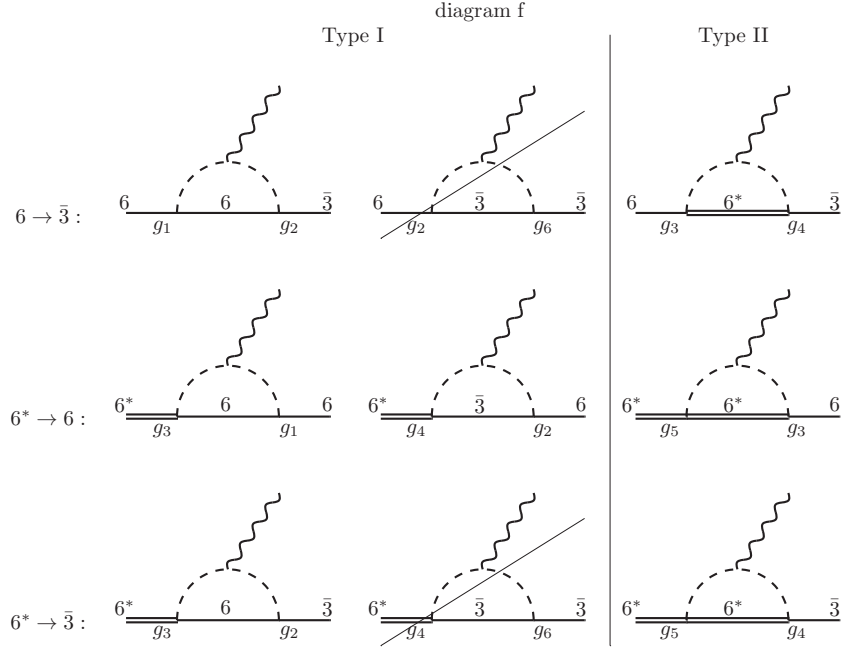}
\end{center}
  \caption{The flavor structures of loop-f.}\label{Fig.4}
\end{figure}

Diagram f can be classified into two different types according to
the Lorentz structure in the loop integrals, which are displayed in
Fig. \ref{Fig.4}. For $g_6=0$, the contributions from two of these
Feynman diagrams in Fig. \ref{Fig.4} vanish.

\subsubsection{spin-$\frac{1}{2}$ to $\frac{1}{2}$ diagram}

For the $6\to \bar 3 $ EM decays, the decay amplitude from the
chiral loops can be written as
\[i\mathcal{M}^{\textrm{loop-f}}=i\mathcal{M}_{\textrm{I}}+i\mathcal{M}_{\textrm{II}}=\epsilon^{\mu\nu\alpha\beta}v_\alpha\bar u^{s_A}_iS_\beta u^{s_B}_j(l_\mu\epsilon_\nu-l_\nu\epsilon_\mu)\left[\frac{i}{2}b_{\textrm{I}}-\frac{2}{3}ib_{\textrm{II}}\right]\]
The function
$\left[\frac{i}{2}b_{\textrm{I}}-\frac{2}{3}ib_{\textrm{II}}\right]$
contains all the coefficients involving $g_i$ etc and the loop
integrals. The loop-f Lorentz structure is the same as that of the
$O(p^2)$ tree diagram. The divergence of the chiral loops can be
absorbed by the $O(p^2)$ LECs. We list the functions of the three
channels below.
\begin{eqnarray*}
\left.\frac{i}{2}b_{\textrm{I}}-\frac{2}{3}ib_{\textrm{II}}\right|_{\Lambda_c^+\leftarrow\Sigma_c^+}
&=&g_3 g_4 \mathit{e} \left(\frac{\gamma _3\left(m_K,-\delta
   _2,\ell\right)}{6 \sqrt{2} F_0^2}+\frac{\gamma
   _9\left(m_K,-\delta _2,\ell\right)}{3 \sqrt{2}
   F_0^2}+\frac{\sqrt{2} \gamma _3\left(m_{\pi },-\delta
   _2,\ell\right)}{3 F_0^2}+\frac{2 \sqrt{2} \gamma
   _9\left(m_{\pi },-\delta _2,\ell\right)}{3
   F_0^2}\right)\\
   &&+g_1 g_2 \mathit{e} \left(\frac{\gamma
   _3\left(m_K,0,\ell\right)}{2 \sqrt{2} F_0^2}+\frac{\gamma
   _9\left(m_K,0,\ell\right)}{\sqrt{2} F_0^2}+\frac{\sqrt{2}
   \gamma _3\left(m_{\pi },0,\ell\right)}{F_0^2}+\frac{2
   \sqrt{2} \gamma _9\left(m_{\pi
   },0,\ell\right)}{F_0^2}\right)\\
\left.\frac{i}{2}b_{\textrm{I}}-\frac{2}{3}ib_{\textrm{II}}\right|_{\Xi_c^+\leftarrow\Xi_c'^+}
&=&g_3 g_4 \mathit{e} \left(\frac{\gamma _3\left(m_K,-\delta
   _2,\ell\right)}{6 \sqrt{2} F_0^2}+\frac{\gamma
   _9\left(m_K,-\delta _2,\ell\right)}{3 \sqrt{2}
   F_0^2}+\frac{\sqrt{2} \gamma _3\left(m_{\pi },-\delta
   _2,\ell\right)}{3 F_0^2}+\frac{2 \sqrt{2} \gamma
   _9\left(m_{\pi },-\delta _2,\ell\right)}{3
   F_0^2}\right)\\
   &&+g_1 g_2 \mathit{e} \left(\frac{\gamma
   _3\left(m_K,0,\ell\right)}{2 \sqrt{2} F_0^2}+\frac{\gamma
   _9\left(m_K,0,\ell\right)}{\sqrt{2} F_0^2}+\frac{\sqrt{2}
   \gamma _3\left(m_{\pi },0,\ell\right)}{F_0^2}+\frac{2
   \sqrt{2} \gamma _9\left(m_{\pi
   },0,\ell\right)}{F_0^2}\right)\\
\left.\frac{i}{2}b_{\textrm{I}}-\frac{2}{3}ib_{\textrm{II}}\right|_{\Xi_c^0\leftarrow\Xi_c'^0}
&=&g_3 g_4 \mathit{e} \left(\frac{\gamma _3\left(m_K,-\delta
   _2,\ell\right)}{6 \sqrt{2} F_0^2}+\frac{\gamma
   _9\left(m_K,-\delta _2,\ell\right)}{3 \sqrt{2}
   F_0^2}-\frac{\gamma _3\left(m_{\pi },-\delta
   _2,\ell\right)}{6 \sqrt{2} F_0^2}-\frac{\gamma
   _9\left(m_{\pi },-\delta _2,\ell\right)}{3 \sqrt{2}
   F_0^2}\right)\\
   &&+g_1 g_2 \mathit{e} \left(\frac{\gamma
   _3\left(m_K,0,\ell\right)}{2 \sqrt{2} F_0^2}+\frac{\gamma
   _9\left(m_K,0,\ell\right)}{\sqrt{2} F_0^2}-\frac{\gamma
   _3\left(m_{\pi },0,\ell\right)}{2 \sqrt{2}
   F_0^2}-\frac{\gamma _9\left(m_{\pi
   },0,\ell\right)}{\sqrt{2} F_0^2}\right)
\end{eqnarray*}
The baryon mass splitting and the meson type in the loop can be read
from the functions directly. The function $\gamma_i(m,\delta,\ell)$
arises from the loop integral, which are defined in Appendix
\ref{app.2}.

\subsubsection{spin-$\frac{3}{2}$ to $\frac{1}{2}$ diagram}

For the spin-$\frac{3}{2}$ to $\frac{1}{2}$ EM decays, the Lorentz
structure of the loop-f amplitude is different from the structure of
the $O(p^2)$ tree level one, but is the same as that of the $O(p^3)$
tree level one. Both type-I and II loop-f amplitudes have two
structures, which are marked with the subscript 1, 2.
\begin{eqnarray*}
i\mathcal{M}^{\textrm{loop-f}}_{\textrm{1}}&=&i\mathcal{M}_{\textrm{I1}}+i\mathcal{M}_{\textrm{II1}}=\bar u^{s_A}S\cdot l\epsilon\cdot u^{s_B}[2b_{\textrm{I}1}+2b_{\textrm{II}1}]\\
i\mathcal{M}^{\textrm{loop-f}}_{\textrm{2}}&=&i\mathcal{M}_{\textrm{I2}}+i\mathcal{M}_{\textrm{II2}}=\bar
u^{s_A}S\cdot\epsilon l\cdot
u^{s_B}[2b_{\textrm{I}2}+2b_{\textrm{II}2}]
\end{eqnarray*}

The function $\left[2b_{\textrm{I}}+ 2b_{\textrm{II}}\right]$
contains all the coefficients involving the coupling constants and
the loop integrals, too. For the six EM decays between the 6 and
$6^*$-dimensional baryons, we have
\begin{eqnarray*}
\left.2b_{\textrm{I}1}+2b_{\textrm{II}1}\right|_{\Sigma_c^{++}\leftarrow\Sigma_c^{*++}}
&=&g_1 g_3 \mathit{e} \left(-\frac{i \gamma _3\left(m_K,\delta
   _2,\ell\right)}{F_0^2}-\frac{i \gamma _9\left(m_K,\delta
   _2,\ell\right)}{F_0^2}-\frac{i \gamma _3\left(m_{\pi
   },\delta _2,\ell\right)}{F_0^2}-\frac{i \gamma
   _9\left(m_{\pi },\delta _2,\ell\right)}{F_0^2}\right)\\
   &&+g_2
   g_4 \mathit{e} \left(-\frac{2 i \gamma _3\left(m_K,\delta
   _3,\ell\right)}{F_0^2}-\frac{2 i \gamma _9\left(m_K,\delta
   _3,\ell\right)}{F_0^2}-\frac{2 i \gamma _3\left(m_{\pi
   },\delta _3,\ell\right)}{F_0^2}-\frac{2 i \gamma
   _9\left(m_{\pi },\delta _3,\ell\right)}{F_0^2}\right)\\
   &&+g_3
   g_5 \mathit{e} \left(\frac{2 i \gamma
   _3\left(m_K,0,\ell\right)}{3 F_0^2}-\frac{i \gamma
   _8\left(m_K,0,\ell\right)}{3 F_0^2}+\frac{2 i \gamma
   _3\left(m_{\pi },0,\ell\right)}{3 F_0^2}-\frac{i \gamma
   _8\left(m_{\pi },0,\ell\right)}{3 F_0^2}\right)\\
\left.2b_{\textrm{I}2}+2b_{\textrm{II}2}\right|_{\Sigma_c^{++}\leftarrow\Sigma_c^{*++}}
&=&g_1 g_3 \mathit{e} \left(-\frac{i \gamma _8\left(m_K,\delta
   _2,\ell\right)}{F_0^2}-\frac{i \gamma _8\left(m_{\pi
   },\delta _2,\ell\right)}{F_0^2}\right)+g_2 g_4 \mathit{e}
   \left(-\frac{2 i \gamma _8\left(m_K,\delta
   _3,\ell\right)}{F_0^2}-\frac{2 i \gamma _8\left(m_{\pi
   },\delta _3,\ell\right)}{F_0^2}\right)\\
   &&+g_3 g_5 \mathit{e}
   \left(-\frac{i \gamma
   _3\left(m_K,0,\ell\right)}{F_0^2}-\frac{i \gamma
   _8\left(m_K,0,\ell\right)}{3 F_0^2}-\frac{i \gamma
   _3\left(m_{\pi },0,\ell\right)}{F_0^2}-\frac{i \gamma
   _8\left(m_{\pi },0,\ell\right)}{3 F_0^2}\right)\\
\left.2b_{\textrm{I}1}+2b_{\textrm{II}1}\right|_{\Sigma_c^+\leftarrow\Sigma_c^{*+}}
&=&g_1 g_3 \mathit{e} \left(-\frac{i \gamma _3\left(m_K,\delta
   _2,\ell\right)}{2 F_0^2}-\frac{i \gamma _9\left(m_K,\delta
   _2,\ell\right)}{2 F_0^2}\right)+g_2 g_4 \mathit{e}
   \left(-\frac{i \gamma _3\left(m_K,\delta
   _3,\ell\right)}{F_0^2}-\frac{i \gamma _9\left(m_K,\delta
   _3,\ell\right)}{F_0^2}\right)\\
   &&+g_3 g_5 \mathit{e}
   \left(\frac{i \gamma _3\left(m_K,0,\ell\right)}{3
   F_0^2}-\frac{i \gamma _8\left(m_K,0,\ell\right)}{6
   F_0^2}\right)\\
\left.2b_{\textrm{I}2}+2b_{\textrm{II}2}\right|_{\Sigma_c^+\leftarrow\Sigma_c^{*+}}
&=&-\frac{i g_1 g_3 \mathit{e} \gamma _8\left(m_K,\delta
   _2,\ell\right)}{2 F_0^2}-\frac{i g_2 g_4 \mathit{e} \gamma
   _8\left(m_K,\delta _3,\ell\right)}{F_0^2}+g_3 g_5
   \mathit{e} \left(-\frac{i \gamma
   _3\left(m_K,0,\ell\right)}{2 F_0^2}-\frac{i \gamma
   _8\left(m_K,0,\ell\right)}{6 F_0^2}\right)\\
\left.2b_{\textrm{I}1}+2b_{\textrm{II}1}\right|_{\Sigma_c^0\leftarrow\Sigma_c^{*0}}
&=&g_1 g_3 \mathit{e} \left(\frac{i \gamma _3\left(m_{\pi },\delta
   _2,\ell\right)}{F_0^2}+\frac{i \gamma _9\left(m_{\pi
   },\delta _2,\ell\right)}{F_0^2}\right)+g_2 g_4 \mathit{e}
   \left(\frac{2 i \gamma _3\left(m_{\pi },\delta
   _3,\ell\right)}{F_0^2}+\frac{2 i \gamma _9\left(m_{\pi
   },\delta _3,\ell\right)}{F_0^2}\right)\\
   &&+g_3 g_5 \mathit{e}
   \left(\frac{i \gamma _8\left(m_{\pi },0,\ell\right)}{3
   F_0^2}-\frac{2 i \gamma _3\left(m_{\pi },0,\ell\right)}{3
   F_0^2}\right)\\
\left.2b_{\textrm{I}2}+2b_{\textrm{II}2}\right|_{\Sigma_c^0\leftarrow\Sigma_c^{*0}}
&=&\frac{i g_1 g_3 \mathit{e} \gamma _8\left(m_{\pi },\delta
   _2,\ell\right)}{F_0^2}+\frac{2 i g_2 g_4 \mathit{e} \gamma
   _8\left(m_{\pi },\delta _3,\ell\right)}{F_0^2}+g_3 g_5
   \mathit{e} \left(\frac{i \gamma _3\left(m_{\pi
   },0,\ell\right)}{F_0^2}+\frac{i \gamma _8\left(m_{\pi
   },0,\ell\right)}{3 F_0^2}\right)\\
\left.2b_{\textrm{I}1}+2b_{\textrm{II}1}\right|_{\Xi_c'^+\leftarrow{\Xi_c^*}'^+}
&=&g_1 g_3 \mathit{e} \left(-\frac{i \gamma _3\left(m_{\pi },\delta
   _2,\ell\right)}{2 F_0^2}-\frac{i \gamma _9\left(m_{\pi
   },\delta _2,\ell\right)}{2 F_0^2}\right)+g_2 g_4 \mathit{e}
   \left(-\frac{i \gamma _3\left(m_{\pi },\delta
   _3,\ell\right)}{F_0^2}-\frac{i \gamma _9\left(m_{\pi
   },\delta _3,\ell\right)}{F_0^2}\right)\\
   &&+g_3 g_5 \mathit{e}
   \left(\frac{i \gamma _3\left(m_{\pi },0,\ell\right)}{3
   F_0^2}-\frac{i \gamma _8\left(m_{\pi },0,\ell\right)}{6
   F_0^2}\right)\\
\left.2b_{\textrm{I}2}+2b_{\textrm{II}2}\right|_{\Xi_c'^+\leftarrow{\Xi_c^*}'^+}
&=&-\frac{i g_1 g_3 \mathit{e} \gamma _8\left(m_{\pi },\delta
   _2,\ell\right)}{2 F_0^2}-\frac{i g_2 g_4 \mathit{e} \gamma
   _8\left(m_{\pi },\delta _3,\ell\right)}{F_0^2}+g_3 g_5
   \mathit{e} \left(-\frac{i \gamma _3\left(m_{\pi
   },0,\ell\right)}{2 F_0^2}-\frac{i \gamma _8\left(m_{\pi
   },0,\ell\right)}{6 F_0^2}\right)\\
\left.2b_{\textrm{I}1}+2b_{\textrm{II}1}\right|_{\Xi_c'^0\leftarrow{\Xi_c^*}'^0}
&=&g_1 g_3 \mathit{e} \left(\frac{i \gamma _3\left(m_K,\delta
   _2,\ell\right)}{2 F_0^2}+\frac{i \gamma _9\left(m_K,\delta
   _2,\ell\right)}{2 F_0^2}+\frac{i \gamma _3\left(m_{\pi
   },\delta _2,\ell\right)}{2 F_0^2}+\frac{i \gamma
   _9\left(m_{\pi },\delta _2,\ell\right)}{2 F_0^2}\right)\\
   &&+g_2
   g_4 \mathit{e} \left(\frac{i \gamma _3\left(m_K,\delta
   _3,\ell\right)}{F_0^2}+\frac{i \gamma _9\left(m_K,\delta
   _3,\ell\right)}{F_0^2}+\frac{i \gamma _3\left(m_{\pi
   },\delta _3,\ell\right)}{F_0^2}+\frac{i \gamma
   _9\left(m_{\pi },\delta _3,\ell\right)}{F_0^2}\right)\\
   &&+g_3
   g_5 \mathit{e} \left(-\frac{i \gamma
   _3\left(m_K,0,\ell\right)}{3 F_0^2}+\frac{i \gamma
   _8\left(m_K,0,\ell\right)}{6 F_0^2}-\frac{i \gamma
   _3\left(m_{\pi },0,\ell\right)}{3 F_0^2}+\frac{i \gamma
   _8\left(m_{\pi },0,\ell\right)}{6 F_0^2}\right)\\
\left.2b_{\textrm{I}2}+2b_{\textrm{II}2}\right|_{\Xi_c'^0\leftarrow{\Xi_c^*}'^0}
&=&g_1 g_3 \mathit{e} \left(\frac{i \gamma _8\left(m_K,\delta
   _2,\ell\right)}{2 F_0^2}+\frac{i \gamma _8\left(m_{\pi
   },\delta _2,\ell\right)}{2 F_0^2}\right)+g_2 g_4 \mathit{e}
   \left(\frac{i \gamma _8\left(m_K,\delta
   _3,\ell\right)}{F_0^2}+\frac{i \gamma _8\left(m_{\pi
   },\delta _3,\ell\right)}{F_0^2}\right)\\
   &&+g_3 g_5 \mathit{e}
   \left(\frac{i \gamma _3\left(m_K,0,\ell\right)}{2
   F_0^2}+\frac{i \gamma _8\left(m_K,0,\ell\right)}{6
   F_0^2}+\frac{i \gamma _3\left(m_{\pi },0,\ell\right)}{2
   F_0^2}+\frac{i \gamma _8\left(m_{\pi },0,\ell\right)}{6
   F_0^2}\right)\\
\left.2b_{\textrm{I}1}+2b_{\textrm{II}1}\right|_{\Omega_c^0\leftarrow\Omega_c^{*0}}
&=&g_1 g_3 \mathit{e} \left(\frac{i \gamma _3\left(m_K,\delta
   _2,\ell\right)}{F_0^2}+\frac{i \gamma _9\left(m_K,\delta
   _2,\ell\right)}{F_0^2}\right)+g_2 g_4 \mathit{e}
   \left(\frac{2 i \gamma _3\left(m_K,\delta
   _3,\ell\right)}{F_0^2}+\frac{2 i \gamma _9\left(m_K,\delta
   _3,\ell\right)}{F_0^2}\right)\\
   &&+g_3 g_5 \mathit{e}
   \left(\frac{i \gamma _8\left(m_K,0,\ell\right)}{3
   F_0^2}-\frac{2 i \gamma _3\left(m_K,0,\ell\right)}{3
   F_0^2}\right)\\
\left.2b_{\textrm{I}2}+2b_{\textrm{II}2}\right|_{\Omega_c^0\leftarrow\Omega_c^{*0}}
&=&\frac{i g_1 g_3 \mathit{e} \gamma _8\left(m_K,\delta
   _2,\ell\right)}{F_0^2}+\frac{2 i g_2 g_4 \mathit{e} \gamma
   _8\left(m_K,\delta _3,\ell\right)}{F_0^2}+g_3 g_5
   \mathit{e} \left(\frac{i \gamma
   _3\left(m_K,0,\ell\right)}{F_0^2}+\frac{i \gamma
   _8\left(m_K,0,\ell\right)}{3 F_0^2}\right)\\
\end{eqnarray*}

For the three EM decays between the $\bar3$ and $6^*$-dimensional
baryons, we have
\begin{eqnarray*}
\left.2b_{\textrm{I}1}+2b_{\textrm{II}1}\right|_{\Lambda_c^+\leftarrow\Sigma_c^{*+}}
&=&g_2 g_3 \mathit{e} \left(-\frac{i \gamma _3\left(m_K,\delta
   _2,\ell\right)}{\sqrt{2} F_0^2}-\frac{i \gamma
   _9\left(m_K,\delta _2,\ell\right)}{\sqrt{2} F_0^2}-\frac{2
   i \sqrt{2} \gamma _3\left(m_{\pi },\delta
   _2,\ell\right)}{F_0^2}-\frac{2 i \sqrt{2} \gamma
   _9\left(m_{\pi },\delta _2,\ell\right)}{F_0^2}\right)\\
   &&+g_4
   g_5 \mathit{e} \left(\frac{i \sqrt{2} \gamma
   _3\left(m_K,0,\ell\right)}{3 F_0^2}-\frac{i \gamma
   _8\left(m_K,0,\ell\right)}{3 \sqrt{2} F_0^2}+\frac{4 i
   \sqrt{2} \gamma _3\left(m_{\pi },0,\ell\right)}{3
   F_0^2}-\frac{2 i \sqrt{2} \gamma _8\left(m_{\pi
   },0,\ell\right)}{3 F_0^2}\right)\\
\left.2b_{\textrm{I}2}+2b_{\textrm{II}2}\right|_{\Lambda_c^+\leftarrow\Sigma_c^{*+}}
&=&g_2 g_3 \mathit{e} \left(-\frac{i \gamma _8\left(m_K,\delta
   _2,\ell\right)}{\sqrt{2} F_0^2}-\frac{2 i \sqrt{2} \gamma
   _8\left(m_{\pi },\delta _2,\ell\right)}{F_0^2}\right)\\
   &&+g_4
   g_5 \mathit{e} \left(-\frac{i \gamma
   _3\left(m_K,0,\ell\right)}{\sqrt{2} F_0^2}-\frac{i \gamma
   _8\left(m_K,0,\ell\right)}{3 \sqrt{2} F_0^2}-\frac{2 i
   \sqrt{2} \gamma _3\left(m_{\pi
   },0,\ell\right)}{F_0^2}-\frac{2 i \sqrt{2} \gamma
   _8\left(m_{\pi },0,\ell\right)}{3 F_0^2}\right)\\
\left.2b_{\textrm{I}1}+2b_{\textrm{II}1}\right|_{\Xi_c^+\leftarrow{\Xi_c^*}'^+}
&=&g_2 g_3 \mathit{e} \left(-\frac{2 i \sqrt{2} \gamma
   _3\left(m_K,\delta _2,\ell\right)}{F_0^2}-\frac{2 i
   \sqrt{2} \gamma _9\left(m_K,\delta
   _2,\ell\right)}{F_0^2}-\frac{i \gamma _3\left(m_{\pi
   },\delta _2,\ell\right)}{\sqrt{2} F_0^2}-\frac{i \gamma
   _9\left(m_{\pi },\delta _2,\ell\right)}{\sqrt{2}
   F_0^2}\right)\\
   &&+g_4 g_5 \mathit{e} \left(\frac{4 i \sqrt{2} \gamma
   _3\left(m_K,0,\ell\right)}{3 F_0^2}-\frac{2 i \sqrt{2}
   \gamma _8\left(m_K,0,\ell\right)}{3 F_0^2}+\frac{i \sqrt{2}
   \gamma _3\left(m_{\pi },0,\ell\right)}{3 F_0^2}-\frac{i
   \gamma _8\left(m_{\pi },0,\ell\right)}{3 \sqrt{2}
   F_0^2}\right)\\
\left.2b_{\textrm{I}2}+2b_{\textrm{II}2}\right|_{\Xi_c^+\leftarrow{\Xi_c^*}'^+}
&=&g_2 g_3 \mathit{e} \left(-\frac{2 i \sqrt{2} \gamma
   _8\left(m_K,\delta _2,\ell\right)}{F_0^2}-\frac{i \gamma
   _8\left(m_{\pi },\delta _2,\ell\right)}{\sqrt{2}
   F_0^2}\right)\\
   &&+g_4 g_5 \mathit{e} \left(-\frac{2 i \sqrt{2}
   \gamma _3\left(m_K,0,\ell\right)}{F_0^2}-\frac{2 i \sqrt{2}
   \gamma _8\left(m_K,0,\ell\right)}{3 F_0^2}-\frac{i \gamma
   _3\left(m_{\pi },0,\ell\right)}{\sqrt{2} F_0^2}-\frac{i
   \gamma _8\left(m_{\pi },0,\ell\right)}{3 \sqrt{2}
   F_0^2}\right)\\
\left.2b_{\textrm{I}1}+2b_{\textrm{II}1}\right|_{\Xi_c^0\leftarrow{\Xi_c^*}'^0}
&=&g_2 g_3 \mathit{e} \left(-\frac{i \gamma _3\left(m_K,\delta
   _2,\ell\right)}{\sqrt{2} F_0^2}-\frac{i \gamma
   _9\left(m_K,\delta _2,\ell\right)}{\sqrt{2} F_0^2}+\frac{i
   \gamma _3\left(m_{\pi },\delta _2,\ell\right)}{\sqrt{2}
   F_0^2}+\frac{i \gamma _9\left(m_{\pi },\delta
   _2,\ell\right)}{\sqrt{2} F_0^2}\right)\\
   &&+g_4 g_5 \mathit{e}
   \left(\frac{i \sqrt{2} \gamma _3\left(m_K,0,\ell\right)}{3
   F_0^2}-\frac{i \gamma _8\left(m_K,0,\ell\right)}{3 \sqrt{2}
   F_0^2}-\frac{i \sqrt{2} \gamma _3\left(m_{\pi
   },0,\ell\right)}{3 F_0^2}+\frac{i \gamma _8\left(m_{\pi
   },0,\ell\right)}{3 \sqrt{2} F_0^2}\right)\\
\left.2b_{\textrm{I}2}+2b_{\textrm{II}2}\right|_{\Xi_c^0\leftarrow{\Xi_c^*}'^0}
&=&g_2 g_3 \mathit{e} \left(\frac{i \gamma _8\left(m_{\pi },\delta
   _2,\ell\right)}{\sqrt{2} F_0^2}-\frac{i \gamma
   _8\left(m_K,\delta _2,\ell\right)}{\sqrt{2}
   F_0^2}\right)\\
   &&+g_4 g_5 \mathit{e} \left(-\frac{i \gamma
   _3\left(m_K,0,\ell\right)}{\sqrt{2} F_0^2}-\frac{i \gamma
   _8\left(m_K,0,\ell\right)}{3 \sqrt{2} F_0^2}+\frac{i \gamma
   _3\left(m_{\pi },0,\ell\right)}{\sqrt{2} F_0^2}+\frac{i
   \gamma _8\left(m_{\pi },0,\ell\right)}{3 \sqrt{2}
   F_0^2}\right)\\
\end{eqnarray*}

\section{The determination of the low energy constnats (LECs)}\label{sec4}

In principle, the low energy constants (LECs) in the chiral
Lagrangians should be extracted through fitting to the experimental
data. However, the available information on the heavy baryon system
is still rather poor, which renders the extraction of the low energy
constants $f,h$ nearly impossible. We try to estimate the LECs using
the quark model.

At the quark level, the flavor wave functions of the spin
$\frac{1}{2}$ baryons can be written as follows.
\begin{equation}\label{eq}
\left\{
\begin{aligned}
&\Lambda_c^+=\frac{1}{\sqrt{2}}(ud-du)c\\
&\Xi_c^+=\frac{1}{\sqrt{2}}(us-su)c\\
&\Xi_c^0=\frac{1}{\sqrt{2}}(ds-sd)c\\
\end{aligned}
\right. \qquad \left\{
\begin{aligned}
&\Sigma_c^{++}=uuc\\
&\Sigma_c^+=\frac{1}{\sqrt{2}}(ud+du)c\\
&\Sigma_c^0=ddc\\
&\Xi_c'^+=\frac{1}{\sqrt{2}}(us+su)c\\
&\Xi_c'^0=\frac{1}{\sqrt{2}}(ds+sd)c\\
&\Omega_c^0=ssc\\
\end{aligned}
\right.
\end{equation}
The flavor wave functions of the spin-$\frac{3}{2}$ baryons are
similar.

The spin wave functions of the heavy baryons are
\begin{equation}\label{eq3.2}
\left\{
\begin{aligned}
&\chi_{\frac{1}{2}\frac{1}{2}}=\frac{1}{\sqrt{2}}(\chi_{\frac{1}{2}}\chi_{-\frac{1}{2}}\chi_{\frac{1}{2}}-\chi_{-\frac{1}{2}}\chi_{\frac{1}{2}}\chi_{\frac{1}{2}})\\
&\chi_{\frac{1}{2}-\frac{1}{2}}=\frac{1}{\sqrt{2}}(\chi_{\frac{1}{2}}\chi_{-\frac{1}{2}}\chi_{-\frac{1}{2}}-\chi_{-\frac{1}{2}}\chi_{\frac{1}{2}}\chi_{-\frac{1}{2}})\\
&\chi_{\frac{1}{2}\frac{1}{2}}=\sqrt{\frac{2}{3}}\chi_{\frac{1}{2}}\chi_{\frac{1}{2}}\chi_{-\frac{1}{2}}-\frac{1}{\sqrt{6}}(\chi_{\frac{1}{2}}\chi_{-\frac{1}{2}}\chi_{\frac{1}{2}}+\chi_{-\frac{1}{2}}\chi_{\frac{1}{2}}\chi_{\frac{1}{2}})\\
&\chi_{\frac{1}{2}-\frac{1}{2}}=\frac{1}{\sqrt{6}}(\chi_{\frac{1}{2}}\chi_{-\frac{1}{2}}\chi_{-\frac{1}{2}}+\chi_{-\frac{1}{2}}\chi_{\frac{1}{2}}\chi_{-\frac{1}{2}})-\sqrt{\frac{2}{3}}\chi_{-\frac{1}{2}}\chi_{-\frac{1}{2}}\chi_{\frac{1}{2}}\\
&\chi_{\frac{3}{2}\frac{3}{2}}=\chi_{\frac{1}{2}}\chi_{\frac{1}{2}}\chi_{\frac{1}{2}}\\
&\chi_{\frac{3}{2}-\frac{3}{2}}=\chi_{-\frac{1}{2}}\chi_{-\frac{1}{2}}\chi_{-\frac{1}{2}}
\end{aligned}
\right.
\end{equation}

Comparing the matrix element of the same EM decay process at both
the hadron and quark level, one can express the LECs in terms of the
constituent quark masses and electric charge. For example, let's
consider the $\Sigma_c^+\rightarrow\gamma+\Lambda_c^+$ decay
process. At the hadron level, the Lagrangian is
\begin{eqnarray}
\mathcal{L}^{(2)}_{\Sigma_c^+\rightarrow\gamma+\Lambda_c^+}=-\frac{2\sqrt{2}}{\Lambda_\chi}f_2e\bar\Lambda_c^+\sigma^{\mu\nu}\Sigma_c^+F_{\mu\nu}
\end{eqnarray}
In the rest frame of the initial state, we choose the momentum of
photon along the 3rd axis with the polarization along the $y$-axis.
The matrix element of such a decay reads
\begin{eqnarray}
\left\langle\Lambda_c^+,-\frac{1}{2}\right|\mathcal{L}_y\left|\Sigma_c^+,\frac{1}{2}\right\rangle&=&-\frac{4\sqrt{2}}{\Lambda_\chi}f_2e
\end{eqnarray}
where we have dropped the photon momentum and the image imaginary
unit etc. These common factors also appear in the matrix element at
the quark level. On the other hand, we consider the same matrix
element at the quark level. The Gordon decomposition separates the
QED interaction Lagrangian into the electric and magnetic parts. The
light quark magnetic interaction reads
\begin{eqnarray}\label{eq5.6}
\mathcal{L}_{\textrm{Im}}=-\frac{1}{4}e\left(\frac{2}{3m_u}\bar
u\sigma^{\mu\nu}u-\frac{1}{3m_d}\bar
d\sigma^{\mu\nu}d-\frac{1}{3m_s}\bar
s\sigma^{\mu\nu}s\right)F_{\mu\nu}
\end{eqnarray}
For the heavy quarks we have
\[\mathcal{L}_{\textrm{Im},c}=-\frac{1}{4}e\frac{2}{3m_c}\bar c\sigma^{\mu\nu}cF_{\mu\nu},\quad\mathcal{L}_{\textrm{Im},b}=\frac{1}{4}e\frac{1}{3m_b}\bar b\sigma^{\mu\nu}bF_{\mu\nu}\]
respectively. We notice that the heavy quark magnetic interaction
only contributes to the EM transitions between the $6^*$ and
6-dimensional baryons.

The flavor-spin wave functions of $\Lambda_c^+$ and $\Sigma_c^+$ at
the quark level are
\begin{eqnarray}
\left\{
\begin{aligned}
&\left|\Lambda_c^+,-\frac{1}{2}\right\rangle=\frac{1}{\sqrt{2}}(|udc\rangle-|duc\rangle)\otimes\left[\frac{1}{\sqrt{2}}(|\uparrow\downarrow\downarrow\rangle-|\downarrow\uparrow\downarrow\rangle)\right]\\
&\left|\Sigma_c^+,\frac{1}{2}\right\rangle=\frac{1}{\sqrt{2}}(|udc\rangle+|duc\rangle)\otimes\left[\sqrt{\frac{2}{3}}|\uparrow\uparrow\downarrow\rangle-\sqrt{\frac{1}{6}}(|\uparrow\downarrow\uparrow\rangle+|\downarrow\uparrow\uparrow\rangle)\right]
\end{aligned}
\right.
\end{eqnarray}
The magnetic interaction and tensor operator $\sigma^{\mu\nu}$ flips
the spin of the up and down diquark system. The matrix element at
the quark level reads
\begin{eqnarray}
\left\langle\Lambda_c^+,-\frac{1}{2}\right|\mathcal{L}_y\left|\Sigma_c^+,\frac{1}{2}\right\rangle=-\frac{e}{3\sqrt{3}}\left(\frac{1}{m_u}+\frac{1}{2m_d}\right)
\end{eqnarray}

With the commonly used values of the constituent quark masses
\cite{Scadron:2006dy}.
\[m_u=335.5\textrm{MeV},m_d=339.5\textrm{MeV},m_s=510\textrm{MeV},m_c=1550\textrm{MeV},m_b=4730\textrm{MeV},\]
one obtains the value of the LEC $f_2$
\[f_2=0.166\]

The extraction of the other LECs for the $6^*\to 6$ process is more
complicated. Let's take the decay channel
$\Sigma_c^{*+}\rightarrow\Sigma_c^+ +\gamma$ for an example. Both
the $O(p^2)$ and $O(p^3)$ Lagrangians will contribute to the matrix
element at the hadron level.
\[\left\langle\Sigma_c^+,\frac{1}{2}\right|\mathcal{L}^{(2)}\left|\Sigma_c^{*+},\frac{3}{2}\right\rangle=\left(-\frac{4}{\sqrt{2}\Lambda_\chi^2}f_3-\frac{8}{\sqrt{2}\Lambda_\chi^2}\tilde{f}_3\right)eM_{\Sigma_c^{*+}}\]
\[\left\langle\Sigma_c^+,\frac{1}{2}\right|\mathcal{L}^{(3)}\left|\Sigma_c^{*+},\frac{3}{2}\right\rangle=\left(-\frac{4(h_{13}+\tilde{h}_{13})}{\sqrt{2}\Lambda_\chi^3}+\frac{4(h_{23}+\tilde{h}_{23})}{\Lambda_\chi^3}\right)e(M_{\Sigma_c^{*+}}+M_{\Sigma_c^+})(M_{\Sigma_c^{*+}}-M_{\Sigma_c^+})\]
The matrix element at the quark level reads
\begin{eqnarray}
\left\langle\Sigma_c^+,\frac{1}{2}\right|\mathcal{L}\left|\Sigma_c^{*+},\frac{3}{2}\right\rangle
=-\frac{e}{3\sqrt{6}}\left(\frac{1}{m_u}-\frac{1}{2m_d}-\frac{1}{m_c}\right)
\end{eqnarray}
There are similar equations at both the hadron and quark level for
all the other $6^*\to 6$ processes.

At $O(p^2)$, there are only two unknown LECs $f_3$ and $\tilde{f}_3$
while there are more equations. The best fit in the SU(3) flavor
symmetry limit leads to
\[f_3=0.099,\quad\tilde{f}_3=-0.037.\]
In the extraction of the $O(p^3)$ LECs $h_{13}$ etc, we use the
above values as input and take into account the explicit SU(3)
symmetry breaking, i.e., using the experimental values of the heavy
baryon masses and different values for $m_u$ and $m_s$. Solving
these above equations simultaneously, we get the values of the LECs
at $O(p^3)$
\[h_{13}=0.073,\quad h_{23}=0.034,\quad\tilde{h}_{13}=0.035,\quad\tilde{h}_{23}=0.061\]
Similarly, we can extract the LECs for the $6^*\to \bar 3$ process.
\[f_4=0.133,\quad h_{46}=0.056,\quad h_{56}=-0.039\]
All the values of LECs are collected in Table \ref{LECs}.


\begin{table}[h]
\centering \caption{The values of the dimensionless LECs for the
charmed baryon systems from the quark model.}\label{LECs}
\begin{tabular}{ccc}
\Xhline{1.2pt}
                & $O(p^2)$ tree level & $O(p^3)$ tree level\\
\Xhline{1.2pt}
$f_2$           & 0.166   &$\times$ \\
\hline
$f_3$           & 0.099   &$\times$ \\
$\tilde{f}_3$   &-0.037   &$\times$ \\
$f_4$           & 0.133   &$\times$ \\
\Xhline{1.2pt}
$h_{13}$        &$\times$ & 0.073   \\
$h_{23}$        &$\times$ & 0.034   \\
$\tilde{h}_{13}$&$\times$ & 0.035   \\
$\tilde{h}_{23}$&$\times$ & 0.061   \\
\Xhline{1.2pt}
$h_{46}$        &$\times$ & 0.056   \\
$h_{56}$        &$\times$ &-0.039   \\
\Xhline{1.2pt}
\end{tabular}
\end{table}

\section{Numerical Results and Discussions}\label{sec5}

So far, all LECs and the chiral loop corrections have been obtained.
The EM decay width reads
\begin{eqnarray}
\Gamma=\frac{\ell}{8\pi M_A^2}|\mathcal{M}|^2
\end{eqnarray}
We collect the numerical results of the EM decay widths of the
ground state charmed baryons in Table \ref{width} (in unit of keV).
The notation "$\times$" means no contribution in this channel and
"-" means the contribution is strongly suppressed. In Table
\ref{width}, the first three rows, middle six rows and the last
three rows correspond to the $6\to \bar 3$, $6^* \to 6$ and $6^* \to
\bar3$ EM decays respectively.

\begin{table}[h]
\centering
\caption{The EM decay width of the charmed baryons in
unit of keV.}\label{width}
\begin{tabular}{cc|cc|cc|c}
\Xhline{1.2pt}
 & & & & $O(p^5)$ Cross term with & $O(p^6)$ Tree+loop &\\
EM decay process                               &$O(p^4)$ tree level & $O(p^5)$ cross term & $O(p^6)$ tree level & loop contribution & correction &Full width\\
\Xhline{1.2pt}
$\Sigma_c^+\rightarrow\gamma+\Lambda_c^+$      & 149.41 &$\times$&$\times$&$\times$&164.16& 164.16\\
$\Xi_c'^+\rightarrow\gamma+\Xi_c^+$            &  43.26 &$\times$&$\times$&$\times$& 54.31& 54.31\\
$\Xi_c'^0\rightarrow\gamma+\Xi_c^0$            &$\times$&$\times$&$\times$&$\times$&  0.02&  0.02\\
\Xhline{1.2pt}
$\Sigma_c^{*++}\rightarrow\gamma+\Sigma_c^{++}$&  8.21  & 1.46 & 0.13&  2.88 & 0.51&11.60\\
$\Sigma_c^{*+}\rightarrow\gamma+\Sigma_c^{+}$  &  0.35  & 0.23 & 0.08&  0.32 & 0.18& 0.85\\
$\Sigma_c^{*0}\rightarrow\gamma+\Sigma_c^{0}$  &  2.93  &-0.41 & 0.03& -0.03 & 0.02& 2.92\\
${\Xi_c^*}'^+\rightarrow\gamma+\Xi_c'^+$       &  0.48  & 0.34 & 0.13&  0.43 & 0.19& 1.10\\
${\Xi_c^*}'^0\rightarrow\gamma+\Xi_c'^0$       &  3.87  &-0.58 & 0.05& -0.04 & -   & 3.83\\
$\Omega_c^{*0}\rightarrow\gamma+\Omega_c^0$    &  4.82  &-0.75 & 0.06& -0.03 & 0.03& 4.82\\
\Xhline{1.2pt}
$\Sigma_c^{*+}\rightarrow\gamma+\Lambda_c^+$   &756.40  & 18.43& 7.23& 125.74&10.83&892.97\\
${\Xi_c^*}'^+\rightarrow\gamma+\Xi_c^+$        &402.92  &  7.63& 2.33&  94.70& 4.49&502.11\\
${\Xi_c^*}'^0\rightarrow\gamma+\Xi_c^0$        &$\times$&$\times$&$\times$&$\times$&0.36& 0.36\\
\Xhline{1.2pt}
\end{tabular}
\end{table}

The EM decay width is proportional to the cubic of the photon
momentum, which depends on the mass difference of the initial and
final heavy baryons. The larger the mass difference, the larger the
decay width. Therefore the $6^*/6 \to \bar3$ EM decay widths may
reach several hundred keV while the $6^*\to 6$ transition width is
only a few keV.

From Table \ref{width}, the numerical results of the EM decay widths
show very good convergence of the chiral expansion. For the $6 \to
\bar3$ decays, the leading order widths are much larger than the
loop contributions. For the nine spin-$\frac{3}{2}$ to $\frac{1}{2}$
transitions, the leading order amplitude yields the EM decay width
at $O(p^4)$. The cross term from the tree-level $O(p^2)$ and
$O(p^3)$ amplitudes contributes to the decay width at $O(p^5)$ while
the $O(p^3)$ amplitude alone contributes to the widths at $O(p^6)$.
From the second, third and fourth column of Table \ref{width}, the
numerical results of the $O(p^4)$, $O(p^5)$ and $O(p^6)$ corrections
decrease accordingly, consistent with the chiral power. When the
loop contributions are added in the fifth and sixth columns, the
numerical values also maintain the same decreasing trend.


A remarkable feature in Table \ref{width} is the strong suppression
of the $\Xi_c'^0\rightarrow\gamma+\Xi_c^0$ and
${\Xi_c^*}'^0\rightarrow\gamma+\Xi_c^0$ decays. As we have pointed
out before, the exact SU(3) U-spin flavor symmetry forbids the two
neutral EM decay processes $\Xi_c'^0\rightarrow\gamma+\Xi_c^0$ and
${\Xi_c^*}'^0\rightarrow\gamma+\Xi_c^0$. There do not exist
tree-level contributions at $O(p^2)$ and $O(p^3)$. The non-zero
tree-level contribution to the decay amplitude starts at $O(p^4)$
which explicitly breaks the SU(3) flavor symmetry through the
current quark mass matrix. On the other hand, the chiral loop
corrections appear at $O(p^3)$ if we use the physical mass for the
eight Goldstone bosons. In fact, the chiral corrections from the
pion, kaon and eta mesons will cancel exactly at $O(p^3)$ if their
masses and decay constants are the same, which is also required by
the U-spin symmetry. From Table \ref{width}, we do notice that the
numerical values of the loop contributions of these two channels are
indeed much smaller than their counterparts in the same multiplet.

In short summary, we have investigated the electromagnetic decays of
the heavy baryon systems within the heavy baryon chiral perturbation
theory. After introducing the leading order chiral Lagrangians and
constructing the $O(p^2)$ and $O(p^3)$ Lagrangians, we have
systematically calculated the twelve EM decay amplitudes up to
$O(p^3)$. With the help of the quark model, we estimate all the low
energy constants at $O(p^2)$ and $O(p^3)$. The convergence of the
chiral expansion is quite good.

In the future, the EM decay widths of the charmed and bottomed
baryons may be measured at facilities such as LHCb and JPARC. We
sincerely hope our present work will be useful to future
experiments. Moreover, the explicit chiral structures of the heavy
baryon decay amplitudes derived in this work may also be useful to
the possible chiral extrapolations of the future lattice simulations
of these EM decay amplitudes.

\section*{ACKNOWLEDGMENTS}

N.J. is very grateful to L. Ma, G.-J. Wang and Professor W. Z. Deng
for very helpful discussions. This project is supported by the
National Natural Science Foundation of China under Grant No.
11261130311.

\newpage

\section*{Appendix}\label{app}

\section{EM decay widths of the bottomed baryons}

In the same framework, we estimate the EM decay widths of the
bottomed baryons. The chiral Lagrangians of the bottomed baryons are
the same as those for the charmed baryons. All the
coefficients $f^b,h^b$ are collected in Table \ref{CoefficientD-b}. They can
also be determined from the quark model. The value of the LECs are
listed in Table \ref{LECs-b}.

\begin{table}[h]
  \centering
  \caption{The dimensionless LECs for the bottomed baryon systems.}\label{LECs-b}
\begin{tabular}{ccc}
\Xhline{1.2pt}
                & $O(p^2)$ tree level & $O(p^3)$ tree level\\
\Xhline{1.2pt}
$f^b_2$           & 0.166   &$\times$ \\
\hline
$f^b_3$           & 0.047   &$\times$ \\
$\tilde{f}^b_3$   &-0.021   &$\times$ \\
$f^b_4$           & 0.058   &$\times$ \\
\Xhline{1.2pt}
$h^b_{13}$        &$\times$ & 0.369   \\
$h^b_{23}$        &$\times$ &-0.058   \\
$\tilde{h}^b_{13}$&$\times$ &-0.096   \\
$\tilde{h}^b_{23}$&$\times$ & 0.068   \\
\Xhline{1.2pt}
$h^b_{46}$        &$\times$ & 0.087   \\
$h^b_{56}$        &$\times$ &-0.061   \\
\Xhline{1.2pt}
\end{tabular}
\end{table}

At present, the experimental data on the bottomed baryons is scarce.
We choose the average value of the bottomed baryon mass as
$M^b_6=5891.87$MeV \cite{Agashe:2014kda}. We naively assume that
the the mass splitting between 6
and $6^*$-dimensional is inversely proportional to the
heavy quark mass.
\[(M^b_{6^*}-M^b_6)m_b=(M^c_{6^*}-M^c_6)m_c\]
Then we obtain the mass splitting for the bottomed baryons as
\[\delta^b_2=21.97\textrm{MeV}\]
We let $\delta^b_1,\delta^b_3$ take roughly the same values as in
the charmed baryon systems since these splittings arise from the
strong interaction of the light diquark system. The above values
will be used in the calculation of the chiral loop corrections.

Since the EM decay widths are very sensitive to the photon momentum,
we explicitly keep the photon momentum $\ell$ in the decay width
formula which shall be measured by future experiments. The EM decay
widths of the bottomed baryons are listed below in unit of keV while
the photon momentum $\ell$ is in unit of MeV.
\begin{eqnarray*}
\Gamma_{\Sigma_b^0\rightarrow\gamma+\Lambda_b^0}&=&4.15\times10^{-5}\ell^3\\
\Gamma_{\Xi_b'^0\rightarrow\gamma+\Xi_b^0}&=&4.67\times10^{-5}\ell^3\\
\Gamma_{\Xi_b'^-\rightarrow\gamma+\Xi_b^-}&=&2.18\times10^{-8}\ell^3\\
\Gamma_{\Sigma_b^{*+}\rightarrow\gamma+\Sigma_b^+}&=&5.37\times10^{-5}\ell^3+5.74\times10^{-7}\ell^4+3.20\times10^{-9}\ell^5\\
\Gamma_{\Sigma_b^{*0}\rightarrow\gamma+\Sigma_b^0}&=&5.12\times10^{-6}\ell^3+3.21\times10^{-8}\ell^4+2.89\times10^{-10}\ell^5\\
\Gamma_{\Sigma_b^{*-}\rightarrow\gamma+\Sigma_b^-}&=&7.86\times10^{-6}\ell^3+13.10\times10^{-8}\ell^4+14.22\times10^{-10}\ell^5\\
\Gamma_{{\Xi_b^*}'^0\rightarrow\gamma+\Xi_b'^0}&=&5.30\times10^{-6}\ell^3+2.98\times10^{-8}\ell^4+2.14\times10^{-10}\ell^5\\
\Gamma_{{\Xi_b^*}'^-\rightarrow\gamma+\Xi_b'^-}&=&7.86\times10^{-6}\ell^3+13.82\times10^{-8}\ell^4+13.64\times10^{-10}\ell^5\\
\Gamma_{\Omega_b^{*-}\rightarrow\gamma+\Omega_b^-}&=&8.29\times10^{-6}\ell^3+15.59\times10^{-8}\ell^4+26.86\times10^{-10}\ell^5\\
\Gamma_{\Sigma_b^{*0}\rightarrow\gamma+\Lambda_b^0}&=&7.55\times10^{-5}\ell^3-1.85\times10^{-7}\ell^4+2.31\times10^{-10}\ell^5\\
\Gamma_{{\Xi_b^*}'^0\rightarrow\gamma+\Xi_b^0}&=&7.95\times10^{-5}\ell^3-3.13\times10^{-7}\ell^4+2.84\times10^{-10}\ell^5\\
\Gamma_{{\Xi_b^*}'^-\rightarrow\gamma+\Xi_b^-}&=&2.15\times10^{-11}\ell^5
\end{eqnarray*}

We also list the numerical results of the EM decay widths in Table
\ref{width-b} where the masses of the initial and final state bottom
baryons are known experimentally \cite{Agashe:2014kda}.

\begin{table}[h]
\centering \caption{The EM decay widths of the bottomed baryons
decay in unit of keV.}\label{width-b}
\begin{tabular}{cc|cc|c}
\Xhline{1.2pt}
 & & $O(p^5)$ Cross term with & $O(p^6)$ Tree+loop &\\
EM decay process                               & $O(p^4)$ tree level & loop contribution & correction &Full width\\
\Xhline{1.2pt}
$\Sigma_b^0\rightarrow\gamma+\Lambda_b^0$      & 255.90 &$\times$&287.65& 287.65\\
\Xhline{1.2pt}
$\Sigma_b^{*+}\rightarrow\gamma+\Sigma_b^+$    &  0.48  &  0.11 & 0.01& 0.60\\
$\Sigma_b^{*0}\rightarrow\gamma+\Sigma_b^0$    &  0.04  &  0.01 & -   & 0.05\\
$\Sigma_b^{*-}\rightarrow\gamma+\Sigma_b^-$    &  0.06  &  0.02 & -   & 0.08\\
\Xhline{1.2pt}
$\Sigma_b^{*0}\rightarrow\gamma+\Lambda_b^0$   &701.63  & -361.29&94.83&435.17\\
${\Xi_b^*}'^0\rightarrow\gamma+\Xi_b^0$        &281.32  & -168.70&23.34&135.96\\
${\Xi_b^*}'^-\rightarrow\gamma+\Xi_b^-$        &$\times$&$\times$& 1.87& 1.87\\
\Xhline{1.2pt}
\end{tabular}
\end{table}

\section{INTEGRALS AND FUNCTIONS}\label{app.2}

\noindent1. The integral with one meson line
\[I(m)=\frac{m^2}{16\pi^2}\left(R+\ln\frac{m^2}{\mu^2}\right)\]

\noindent2. The integral with one meson line and one baryon line

\begin{eqnarray}\label{eq33}
J(m,\omega)&=&\frac{1}{8\pi^2}\left[\omega(R-1)+\omega\ln\frac{m^2}{\mu^2}+K\right],\omega=v\cdot
k+\delta
\end{eqnarray}

and
\[K=
\left\{
\begin{aligned}
&2\sqrt{\omega^2-m^2}\textrm{arccosh}\frac{\omega}{m}-2i\pi\sqrt{\omega^2-m^2},\quad\omega>m\\
&-2\sqrt{\omega^2-m^2}\textrm{arccosh}\frac{-\omega}{m},\quad\omega<-m\\
&2\sqrt{m^2-\omega^2}\arccos\frac{-\omega}{m},\quad\omega^2<m^2\\
\end{aligned}
\right.
\]

\noindent3. The integral with two meson lines and one baryon line

3.1 scaler integral
\begin{eqnarray}
L&=&\frac{1}{i}\int\frac{d^dq}{(2\pi)^d}\frac{1}{(m^2-(q+l)^2)(m^2-q^2)[v\cdot
q+\omega]},\quad l^2=0
\end{eqnarray}

\begin{eqnarray}
L=\frac{1}{8\pi^2v\cdot l} \left\{
\begin{aligned}
&-\frac{1}{2}\left[\left(\textrm{arccosh}\frac{-\omega+v\cdot l}{m}\right)^2-\left(\textrm{arccosh}\frac{-\omega}{m}\right)^2\right]\\
&\qquad+\pi i\left[\ln\left(\sqrt{(\omega-v\cdot l)^2-m^2}+\omega-v\cdot l\right)-\ln\left(\sqrt{\omega^2-m^2}+\omega\right)\right],\quad\beta>m\\
&-\frac{1}{2}\left[\left(\textrm{arccosh}\frac{-\omega+v\cdot l}{m}\right)^2-\left(\textrm{arccosh}\frac{-\omega}{m}\right)^2\right],\quad\beta<-m\\
&\frac{1}{2}\left[\left(\arccos\frac{-\omega+v\cdot
l}{m}\right)^2-\left(\arccos\frac{-\omega}{m}\right)^2\right],\quad\beta^2<m^2
\end{aligned}
\right.
\end{eqnarray}
where$\beta=\omega-v\cdot l$

3.2 vector integral
\begin{eqnarray}
L_\mu&=&\frac{1}{i}\int\frac{d^dq}{(2\pi)^d}\frac{q_\mu}{(m^2-(q+l)^2)(m^2-q^2)[v\cdot
q+\omega]}=\gamma_1l_\mu+\gamma_2v_\mu
\end{eqnarray}
\begin{eqnarray}
\gamma_1(m,\omega,v\cdot l)&=&\frac{1}{v\cdot l}(I_2(m)-\omega L(m,\omega,v\cdot l))-\frac{1}{2(v\cdot l)^2}[J(m,\beta)-J(m,\omega)]\\
\gamma_2(m,\omega,v\cdot l)&=&\frac{1}{2(v\cdot
l)^2}[J(m,\beta)-J(m,\omega)]
\end{eqnarray}
in the rest frame, $v\cdot l=\ell$, which is the modular of photon
3-momentum.

3.3 Second-order integral

\begin{eqnarray}
L_{\mu\nu}&=&\frac{1}{i}\int\frac{d^dq}{(2\pi)^d}\frac{q_\mu q_\nu}{(m^2-(q+l)^2)(m^2-q^2)[v\cdot q+\omega]}\\
&=&\gamma_3g_{\mu\nu}+\gamma_4l_\mu l_\nu+\gamma_5v_\mu
l_\nu+\gamma_6l_\mu v_\nu+\gamma_7v_\mu v_\nu\nonumber
\end{eqnarray}

\begin{eqnarray}
\gamma_3(m,\omega,\ell )&=&-\frac{1}{2 (d-2) \ell }\{\omega  J(m,\omega )+(\beta -2 \omega +2 \ell ) J(m,\beta )-2 m^2 \ell  L(m,\omega ,\ell )\}\\
\gamma_4(m,\omega,\ell )&=&\frac{1}{2 (d-2) \ell ^3}\{(\omega -d \omega ) J(m,\omega )+J(m,\beta ) (-(d-1) (\beta -2 \omega )-2 \ell )\nonumber\\
&&+2 \ell  \left((d-2) \omega ^2+m^2\right) L(m,\omega ,\ell )-(d-2) \ell  I_2(m) (2 \omega +\ell )\}\\
\gamma_{5,6}(m,\omega,\ell )&=&\frac{1}{2 (d-2) \ell ^2}\{J(m,\beta ) (\beta  (d-1)-d \omega +2 \ell )+\omega  J(m,\omega )-2 m^2 \ell  L(m,\omega ,\ell )\}\\
\gamma_7(m,\omega,\ell )&=&\frac{1}{2 \ell }\{\omega  J(m,\omega
)-\beta  J(m,\beta )\}
\end{eqnarray}
$d$ arises from the dimensional regularization in d-dimensional
space.

3.4 Third-order integral

\begin{eqnarray}
L_{\mu\alpha\beta}&=&\frac{1}{i}\int\frac{d^dq}{(2\pi)^d}\frac{q_\mu q_\alpha q_\beta}{(m^2-(q+l)^2)(m^2-q^2)[v\cdot q+\omega]}\\
&=&\gamma_8g_{\mu\alpha}l_\beta+\gamma_9g_{\mu\beta}l_\alpha+\gamma_{10}g_{\alpha\beta}l_\mu\nonumber\\
&&+\gamma_{11}g_{\mu\alpha}v_\beta+\gamma_{12}g_{\mu\beta}v_\alpha+\gamma_{13}g_{\alpha\beta}v_\mu\nonumber\\
&&+\gamma_{14}l_\mu l_\alpha l_\beta+\gamma_{15}l_\mu l_\alpha v_\beta+\gamma_{16}l_\mu v_\alpha l_\beta+\gamma_{17}v_\mu l_\alpha l_\beta\nonumber\\
&&+\gamma_{18}l_\mu v_\alpha v_\beta+\gamma_{19}v_\mu l_\alpha
v_\beta+\gamma_{20}v_\mu v_\alpha l_\beta+\gamma_{21}v_\mu v_\alpha
v_\beta\nonumber
\end{eqnarray}

For completeness, we list all the functions $\gamma_i$ as follows,
although only $\gamma_8$ and $\gamma_9$ are used in the calculations
of loop-f. Some functions may be useful in the calculation of the
magnetic moment of heavy baryons.

{\small
\begin{eqnarray}
\gamma_{8,9,10}(m,\omega,\ell )&=&\frac{1}{2 (d-2) (d-1) \ell ^2}\{-\left(\beta ^2+(d-2) m^2\right) J(m,\beta )\\
&&+J(m,\omega ) \left((d-2) m^2+\omega  (-d \ell +\omega +\ell )\right)+\ell  \left(2 (d-1) m^2 \left(I_2(m)-L \omega \right)+(d-2) I(m)\right)\}\nonumber\\
\gamma_{11,12,13}(m,\omega,\ell )&=&\frac{1}{2 (d-2) (d-1) \ell }\{J(m,\beta ) \left((d-2) m^2+\beta  (2 (d-1) \ell -\beta (d-2))\right)\\
&&+J(m,\omega ) \left(-(d-2) m^2+(d-2) \omega ^2-2 (d-1) \omega  \ell \right)-(d-2) \ell  I(m)\}\nonumber\\
\gamma_{14}(m,\omega,\ell )&=&\frac{1}{6 (d-2) (d-1) d \ell ^4}\{-3 d J(m,\beta ) \left(\beta ^2 ((d-2) d+3)+(d-2) \left(3 (d-1) \ell ^2+2 m^2\right)\right)\\
&&+3 d J(m,\omega ) \left(2 (d-2) m^2+\omega \left(((d-2) d+3) \omega -3 (d-1)^2 \ell \right)\right)\nonumber\\
&&-6 (d-1) d L \omega  \ell  \left((d-2) \omega ^2+3 m^2\right)+(d-1) \ell  I_2(m) \left(12 (2 d-1) m^2+(d-2) d \left(6 \omega ^2+2 \ell ^2+3 \omega  \ell \right)\right)\nonumber\\
&&+6 (d-2) ((d-1) d+1) \ell  I(m)\}\nonumber\\
\gamma_{15,16,17}(m,\omega,\ell )&=&\frac{1}{2 (d-2) (d-1) \ell ^3}\{2 \left(\ell  \left(2 (d-1) m^2 \left(L \omega -I_2(m)\right)-(d-2) I(m)\right)-\left((d-2) m^2+\omega ^2\right) J(m,\omega )\right)\nonumber\\
&&+J(m,\beta ) \left(2 \beta ^2+(d-2) \left((d-1) \ell ^2+2 m^2\right)+2 \beta  (d-1) \ell \right)\}\\
\gamma_{18,19,20}(m,\omega,\ell )&=&\frac{1}{2 (d-2) (d-1) \ell ^2}\{J(m,\omega ) \left(\left(d^2+d-2\right) \omega  \ell +(d-2) m^2-(d-2) \omega ^2\right)\\
&&+J(m,\beta ) \left(\beta  (\beta  (d-2)-4 (d-1) \ell )-(d-2) m^2\right)+(d-2)^2 (-\ell ) I(m)\}\nonumber\\
\gamma_{21}(m,\omega,\ell )&=&\frac{1}{2 (d-1) \ell }\{\left(\beta
^2 d-m^2\right) J(m,\beta )+\left(m^2-d \omega ^2\right) J(m,\omega
)+d \ell  I(m)\}
\end{eqnarray}
}

\end{document}